\documentclass[12pt,preprint]{aastex}

\def\chandra {{\it Chandra}}
\def\fuse {{\it FUSE}}
\def\sherpa {{\it Sherpa}}
\def\novi {N_{\rm OVI}}
\def\novii {N_{\rm OVII}}
\def\ov {\ion{O}{5}}
\def\ovi {\ion{O}{6}}
\def\ovilv {\ion{O}{6}$_{\rm LV}$}
\def\ovii {\ion{O}{7}}
\def\oviii {\ion{O}{8}}
\def\cv {\ion{C}{5}}
\def\cvi {\ion{C}{6}}
\def\nvi {\ion{N}{6}}
\def\nvii {\ion{N}{7}}
\def\neix {\ion{Ne}{9}}
\def\arxv {\ion{Ar}{15}}
\def\nex {\ion{Ne}{10}}
\def\flu {f_{lu}}
\def\kms{{\rm\,km\,s}^{-1}}

\begin{document}

\title{Probing the Local Group Medium toward Mkn~421 with \chandra\ and \fuse}
\author{Rik J.~Williams\altaffilmark{1}, 
        Smita Mathur\altaffilmark{1},
        Fabrizio Nicastro\altaffilmark{2,3}, 
        Martin Elvis\altaffilmark{2}, 
        Jeremy J.~Drake\altaffilmark{2},  
        Taotao Fang\altaffilmark{4,5}, 
        Fabrizio Fiore\altaffilmark{6},
        Yair Krongold\altaffilmark{2,3}, 
        Q.~Daniel Wang\altaffilmark{7},
        Yangsen Yao\altaffilmark{7}}
\altaffiltext{1}{Department of Astronomy, The Ohio State University, 
                 140 West 18th Avenue, Columbus, OH 43210, USA}
\altaffiltext{2}{Harvard-Smithsonian Center for Astrophysics, Cambridge, MA, 
                 USA}
\altaffiltext{3}{Instituto de Astronom\'ia Universidad Aut\'onomica de
 M\'exico, Apartado Postal 70-264, Ciudad Universitaria, M\'exico, D.F.,
 CP 04510, M\'exico}
\altaffiltext{4}{University of California--Berkeley, Berkeley, CA, USA}
\altaffiltext{5}{\chandra\ Fellow}
\altaffiltext{6}{Osservatorio Astronomico di Roma, Monteporzio, RM, Italy}
\altaffiltext{7}{Department of Astronomy, University of Massachusetts,
                 Amherst, MA  01003, USA}
\email{williams,smita@astronomy.ohio-state.edu; fnicastro@cfa.harvard.edu}
\begin{abstract}

We report the detection of highly--ionized gas at $z\sim 0$ seen in
resonant \mbox{UV} and X-ray absorption lines toward the $z=0.03$ blazar
Mkn~421.  A total of 13 X-ray and 3 UV lines were measured (or
upper limits derived), including three lines in the \ovii\ K-series and
K$\alpha$ transitions from neon, carbon, and nitrogen.  From the three
\ovii\ lines we derive a $2\sigma$ Doppler parameter constraint of 
$24<b<55\kms$.  The \fuse\ spectrum
shows strong Galactic low--velocity \ovi\ 1032\,\AA\ absorption as 
well as a possible
weak \ovi\ high--velocity component (HVC).  The Doppler parameter of the 
low--velocity \ovi\ measured with \fuse\ is $\sim 3\sigma$ higher than
that derived from the \ovii\ line ratios, indicating that the \ovii\ and 
Galactic \ovi\ arise in different phases.  This velocity dispersion,
along with limits on the gas temperature and density from the X-ray line ratios 
(assuming a single phase with collisional ionization equilibrium plus 
photoionization) are
all consistent with an extragalactic absorber.  However, the \ovii\ Doppler
parameter is inconsistent with the high temperature required to produce
the observed
\ovi$_{\rm HVC}$/\ovii\ ratio, implying that the HVC is probably 
not related to the \ovii. 
In addition, the \ovi$_{K\alpha}$ line detected by
\chandra\ implies a column density $\sim 4$ times
higher than the 1032\,\AA\ absorption.  Although an extragalactic absorber
is fully consistent with the measured column density ratios, a Galactic 
origin cannot be ruled out given the uncertainties in the available data.


\end{abstract}


\section{Introduction}
At both low and high redshifts, the intergalactic medium (IGM) is thought
to be the dominant reservoir of baryonic matter in the universe.
Essentially all of the baryons at $z\sim 2$ are accounted for in the 
``forest'' of \ion{H}{1} Lyman--$\alpha$ absorption lines seen against 
background quasars \citep{weinberg97}.  In the local universe, 
numerical simulations predict that most of
the IGM is in the form of a shock--heated ``warm--hot IGM'' (WHIM) with
$T\sim 10^5-10^7$\,K and densities $n_H\sim 10^{-6}-10^{-5}$\,cm$^{-3}$
\citep{cen99,dave01}.  Hydrogen is mostly ionized
at these temperatures, rendering it difficult or impossible to detect.
However, H--,  He--, and Li--like ions of heavier elements can exist under these
conditions, producing a ``forest'' of low--$z$ intervening UV/X-ray absorption
lines in quasar spectra \citep{hellsten98,perna98}.  In particular, \ovi, \ovii, 
and \oviii\ are expected to produce the strongest lines (depending on
the temperature and ionizing background).

The required sensitivity to detect such absorption systems
is now available with the advent of such facilities as the \chandra\ X-ray
observatory, {\it XMM-Newton}, {\it Hubble Space Telescope}, 
and the {\it Far Ultraviolet Spectroscopic 
Explorer} (\fuse).  High signal--to--noise observations along four lines
of sight reveal such intervening absorption systems with temperatures and
densities consistent with the WHIM 
\citep[][Fang et al., in prep]{nicastro05a,nicastro05b,nicastro05c,mathur03,fang02}.  
Additionally,
the total baryonic mass density in these detected filaments is consistent
with the $(2.5\pm 0.4)$\% deficiency between CMB measurements 
\citep{bennett03} and previous density measurements of stars and gas
\citep[see also][and references therein]{nicastro05a}.
  
Since most galaxies are expected to trace the same cosmic overdensities
as the web of IGM filaments, it would be no surprise if the Galaxy itself
resided in such a filament.  Indeed, X-ray spectra of several quasars
show likely $z=0$ \ovii\ absorption, e.g.\ PKS~2155--304 \citep{nicastro02},  
3C~273 \citep{fang03}, and Mkn~279 (Williams, Rodriguez et al., 
in preparation).
The upper limit found for \ovii\ emission implies that \ovii\ \emph{absorption} 
systems most likely arise in an extragalactic medium \citep{rasmussen03}; 
however, the analysis is complicated by $z=0$ \ovi\ absorption 
observed with
\fuse\ along many quasar lines of sight \citep{wakker03}.  The strongest
\ovi\ absorption component, typically within $v_{\rm LSR}=\pm 100\kms$
(hereafter denoted \ovilv), 
is due to gas in the halo and thick disk of the Galaxy \citep{savage03}.
Of the 102 sightlines studied, 84 show at least one additional \ovi\ 
high--velocity component (HVC), typically weaker than the 
\ovilv.

The origin and location of the \ovi\ HVCs is controversial: while some may be 
spatially and
kinematically associated with known Galactic structure (such as 
the Magellanic stream and \ion{H}{1} high--velocity clouds, though this 
association is in many cases uncertain), others appear to be 
completely isolated.  
\citet{sembach03} summarizes the evidence in favor of a Galactic origin for the
\ovi\ HVCs.  In such a scenario, the absorption could arise at the interface
between infalling cooler gas clouds and very hot ($T>10^6$\,K) gas in an 
extended Galactic corona.  The observed gas stripping and orbital deflection
of the Magellanic clouds also lends credence to such a low--density corona.
Furthermore, lower--ionization species such as \ion{Si}{4} and \ion{C}{4}
are seen toward some sightlines.  Thus, there is some evidence for
a collisionally--ionized, high--density Galactic medium.

On the other hand, there is also evidence that the isolated \ovi\ HVCs 
trace the 
Local Group medium and thus are associated with the $z=0$ \ovii\ and \oviii\
X-ray absorption.  The most compelling evidence for this interpretation
is the velocity segregation of the \ovi\ HVCs: \citet{nicastro03} report
that the mean velocity vector of the \ovi\ HVCs is highest in the local 
standard of rest and lowest in the Local Group rest frame, implying an 
extragalactic, Local Group origin.  The presence of lower--ionization lines 
does not
rule out a WHIM origin for the \ovi\ HVCs since the medium is likely
inhomogeneous and may thus consist of phases of different temperatures
and densities \citep[such as the intervening WHIM absorbers 
found by][]{danforth05}.
Furthermore, isolated high--velocity \ovi\ (which is not
associated with \ion{O}{1} and other low--ionization lines) has not
been detected toward high latitude Galactic sources out to 10\,kpc 
\citep{danforth02,howk03,zsargo03}.  The low--ionization line profiles
often differ substantially from the \ovi\ HVCs, implying that they
may arise in different components \citep{collins05}.

The key to understanding the \ovi\ absorption is higher--quality X-ray
data, since the \ovii\ and \oviii\ along high--latitude quasar sightlines are
the most likely to trace the WHIM (i.e., least subject to possible 
contamination by hot Galactic gas).  
X-ray measurements by themselves provide great insight into the properties
of the local WHIM, including its mass and extent, thereby yielding valuable 
constraints for models of galaxy formation and the Local Group
\citep[e.g.,][]{kravtsov02}.  However, knowing whether the \ovi\ HVCs are
also associated with the WHIM provides even stronger diagnostics.  
Unfortunately, existing X-ray data have not been of sufficiently high
quality to resolve this ambiguity: even at large column densities 
($N_i\sim 10^{15}$\,cm$^{-2}$), X-ray
lines are quite weak, and several thousand counts per resolution element
in a typical \chandra\ grating spectrum are required to accurately
measure these lines.  

Through our program of observing nearby blazars
in outburst phases, we have obtained high--quality \chandra\
and \fuse\ spectra of Mkn~421, sufficient to study in 
detail the local WHIM (and Galactic halo/thick disk) absorption.  Here
we report on these observations, and the inferred properties of the
local absorption.

\section{Observations and Data Preparation}
\subsection{\chandra\label{sec_obs}}
A full description of the \chandra\ observations, data reduction,
and continuum fitting can be found in \citet{nicastro05a}; a brief summary
follows.  Mkn~421 was observed during two exceptionally high
outburst phases for 100 ks each as part of our \chandra --AO4 
observing program:
one at $f_{\rm 0.5-2 keV}=1.2\times 10^{-9}$\,erg\,s$^{-1}$\,cm$^{-2}$ 
with the Low Energy Transmission Grating (LETG) combined with the
Advanced CCD Imaging Spectrometer--Spectroscopic \citep[ACIS-S;][]{garmire03}
array, and another
at $f_{\rm 0.5-2 keV}=0.8\times 10^{-9}$\,erg\,s$^{-1}$\,cm$^{-2}$ 
with the High Resolution Camera--Spectroscopic \citep[HRC-S;][]{murray85} 
array and LETG.  Each of these observations
contains $\sim 2500$ counts per resolution element at 21.6\,\AA.  Additionally,
another short observation of Mkn~421 was taken with HRC/LETG
(29 May 2004), providing another 170 counts per resolution element.  These
three spectra were combined over the 10--60\,\AA\ range to improve the 
signal--to--noise ratio (S/N$\sim 55$ at 21\,\AA\ with 0.0125\,\AA\ binning).
The final coadded spectrum of Mkn~421 is one of the best ever taken
with \chandra: it contains
over $10^6$ total counts with $\sim 6000$ counts per resolution element at
21.6\,\AA, providing a $3\sigma$ detection threshold of 
$W_\lambda\sim 2$\,m\AA\ 
($N_{\rm OVII}=8\times 10^{14}\rm{cm}^{-2}$ for an unsaturated line).

Effective area files (ARFs) for each observation were built using 
CIAO\footnote{\url{http://cxc.harvard.edu/ciao/}}
v3.0.2 and CALDB\footnote{\url{http://cxc.harvard.edu/caldb/}} v2.2.6.   
Those pertaining to the ACIS/LETG observations were corrected for the ACIS
quantum efficiency degradation\footnote{See also 
\url{http://cxc.harvard.edu/ciao/why/acisqedeg.html}}
\citep{marshall03}.  For the HRC/LETG observations, the standard
ARFs were used.  Each ARF was then convolved with the relevant standard 
redistribution matrix file (RMF), and the convolved RMFs were
weighted by exposure time, rebinned to the same energy scale,
and averaged to provide a response file for the coadded spectrum.  

Using the CIAO fitting package 
\sherpa\footnote{\url{http://cxc.harvard.edu/sherpa/}}, 
we initially modeled the continuum
of Mkn~421 as a simple power law and a Galactic absorbing
column density of $N_H=1.4\times 10^{20}$\,cm$^{-2}$ \citep{dickey90}, 
excluding the 48--57\,\AA\ HRC chip gap region.  Metal abundances for
the Galactic gas were then artificially adjusted to provide a better fit
around the \ion{O}{1} and \ion{C}{1} K--edges near 23\,\AA\ and 43\,\AA\
respectively.  This is \emph{not} intended to represent actual
changes to the absorber composition, but rather to correct uncertainties
in the instrument calibration.  These adjustments affect the continuum
mostly near the carbon, oxygen, and neon edges, but individual narrow
absorption lines are unaffected.
After this fit there were still some systematic uncertainties in
the best--fit continuum model; these were corrected with 
broad (${\rm FWHM}=0.15-5$\,\AA) Gaussian emission and absorption components 
until the modeled continuum appeared to match the data upon inspection.  
Indeed, the residuals of the spectrum to the final continuum model have a 
nearly Gaussian distribution, with a negative tail indicating the presence
of narrow absorption lines \citep[see][Figure 8]{nicastro05a}.

\subsection{\fuse}

Mkn~421 was also observed with \fuse\ as part of our Director's Discretionary
Time observing program
on 19--21 January 2003 for a total of 62.8\,ks.  An additional 
21.8\,ks observation from 1 December 2000 was also available in the 
archive.  We used the time--tagged, calibrated data from only the LiF1A 
detector, since inclusion of the LiF2B data provides a small ($\sim 20\%$) 
increase in S/N but degrades the overall spectral resolution.\footnote{See the 
\fuse\ Data Analysis Cookbook v1.0, 
\url{http://fuse.pha.jhu.edu/analysis/analysis.html}}
These two observing programs consist of four observations, which in turn
contain a total of 29 individual exposures.
The wavelength scales of each observation's constituent exposures
were cross--correlated and shifted (typically by 1--2 pixels) to account
for slight uncertainties in the wavelength calibration.  The exposures for
each observation were checked for consistency and coadded, weighted
by exposure time.  The resulting four spectra were then cross-correlated
against each other, coadded (with a $\sim 10\%$
downward shift in flux applied to the 2000 observation due to source 
variability), and binned by 5 pixels (0.034\,\AA, or one--half of the nominal
$20\kms$ resolution) providing a S/N of 17 near 1032\,\AA.

To check the absolute wavelength calibration we followed the method of
\citet{wakker03}, using their 4--component fit to the 
\citet{murphy96} Green Bank \ion{H}{1}--21\,cm data as a velocity
reference.  They find four main components of \ion{H}{1} emission
with an $N_H$--weighted average velocity of $-31.7\kms$.  
In the \fuse\ spectrum, the \ion{Si}{2}~$\lambda 1020.699$\,\AA\ and
\ion{Ar}{1}~$\lambda 1048.220$\,\AA\ lines are expected to trace the 
same gas as the 
\ion{H}{1} emission.  Each UV line was fit with two Gaussian
components in \sherpa, giving average velocity offsets of $-30.9$ and 
$-34.9\kms$ respectively.  These agree well with the \ion{H}{1} data,
though the slight difference between the \ion{Ar}{1} and \ion{Si}{2} 
measurements suggest at least a $\sim 4\kms$ intrinsic wavelength uncertainty.

\section{Line Measurements}

\begin{figure}
\plotone{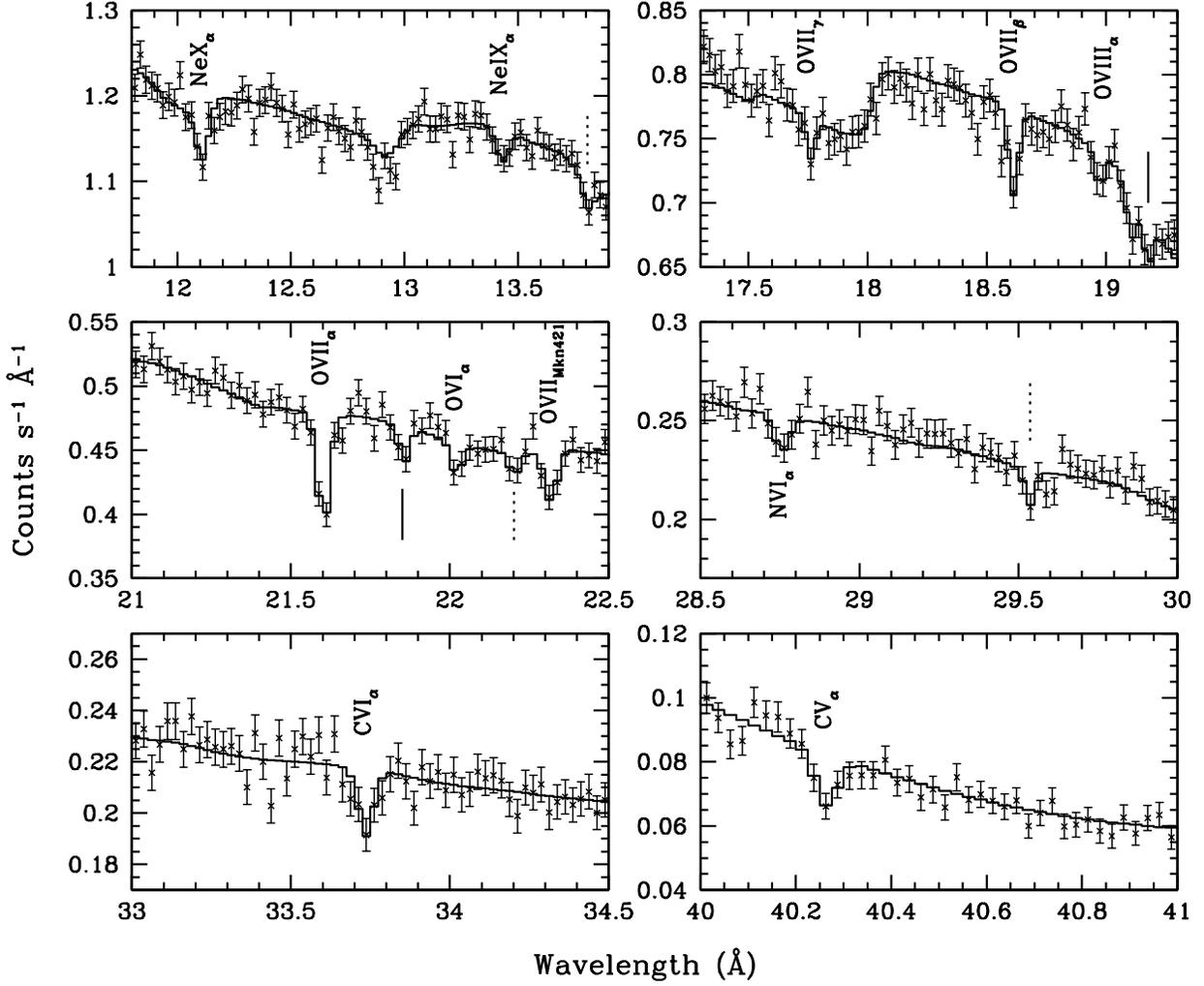}
\caption{Portions of the Mkn~421 \chandra\ LETG spectrum (points) and
the best--fitting model (histogram).  Absorption lines at $z\sim 0$
are labeled, and vertical tick marks indicate absorption from the
$z=0.011$ (solid) and $z=0.027$ (dotted) intervening WHIM filaments
\citep{nicastro05a}. \label{fig_cspec}}
\end{figure}

To find and identify narrow absorption lines in the \chandra\ spectrum of 
Mkn~421,
we visually inspected small (2--5\,\AA) regions of the spectrum, beginning
around the rest wavelength of \ovii$_{K\alpha}$\ (21.602\,\AA) since this
tends to be the strongest $z=0$ X-ray absorption line
\citep[e.g.][]{nicastro02,fang03,chen03}.
Three \ovii$_{K\alpha}$ (Figure~\ref{fig_cspec}) absorption features were
found: one at $z=0$, one at $z=0.011$, and one at $z=0.027$ (with
typical redshift errors of 0.001).  There is also a strong feature which
is $\sim 3\sigma$ from the \ov$_{K\alpha}$ rest wavelength, but is more likely 
\ovii$_{K\alpha}$ at $v\sim +900\kms$ relative to the blazar.  
A close pair of lines consistent with Ly$\alpha$ at this velocity
has been observed \citep{shull96,penton00}, so this may be indicative of an
inflow to Mkn~421 or uncertainty in the blazar redshift 
\citep[based on rather old spectrophotometric measurements by][]{margon78}. 
A weak \ovi$_{K\alpha}$ line is 
seen at the rest wavelength of 22.02\,\AA.  Other regions of the spectrum were 
then searched for lines corresponding to these systems, with particular 
emphasis paid to strong transitions of the most abundant elements (C, N, O, 
and Ne).  All in all there were 13 lines marginally or strongly detected 
at $z\sim 0$ (including the \nvii, \ov, and \arxv\ upper limits), 3 at 
$z=0.011$, and 7 at $z=0.027$.  The latter two systems are the
subject of other papers (Nicastro et al.~2004a,b) and thus will not be
discussed further here.

These 13 $z=0$ X-ray lines were fitted in \sherpa\ with narrow Gaussian features
superposed on the fitted continuum described in \S\ref{sec_obs} (see
Figure~\ref{fig_cspec}).  We are excluding the strong
\ion{O}{1} (23.51\,\AA) line since it arises in the neutral ISM and is not
of interest here, as well as the O$_2$ (23.34\,\AA) absorption since it
coincides with a strong instrumental feature and cannot be accurately measured.
Due to the $\rm{FWHM}=0.04$\,\AA\ ($\sim 600\kms$) LETG resolution the 
lines are all
unresolved, so only the position and equivalent width of each line are
measured.  Errors are calculated using the ``projection'' command in
\sherpa, allowing the overall continuum normalization to vary along with all
parameters for each line.  The resulting line parameter estimates
are presented in Table~\ref{tab_lines}.  The $\sim 0.02$\,\AA\ 
systematic wavelength uncertainty of the 
LETG\footnote{\url{http://cxc.harvard.edu/cal/}} 
is in most cases larger than the
statistical uncertainty of the line centroid; thus, Table~\ref{tab_lines}
lists whichever is greater.  Additionally, 
a meaningful upper error bar on the \cvi\ equivalent width could not be
calculated with \sherpa.  In this case, the FWHM was frozen at the 
instrumental resolution and the error was recalculated; a visual inspection
confirms the new limit to be more reasonable.
Upper limits for the \ov, \nvii, \arxv, and \nex\ lines were calculated with
both the position and FWHM frozen.  

\begin{figure}
\plotone{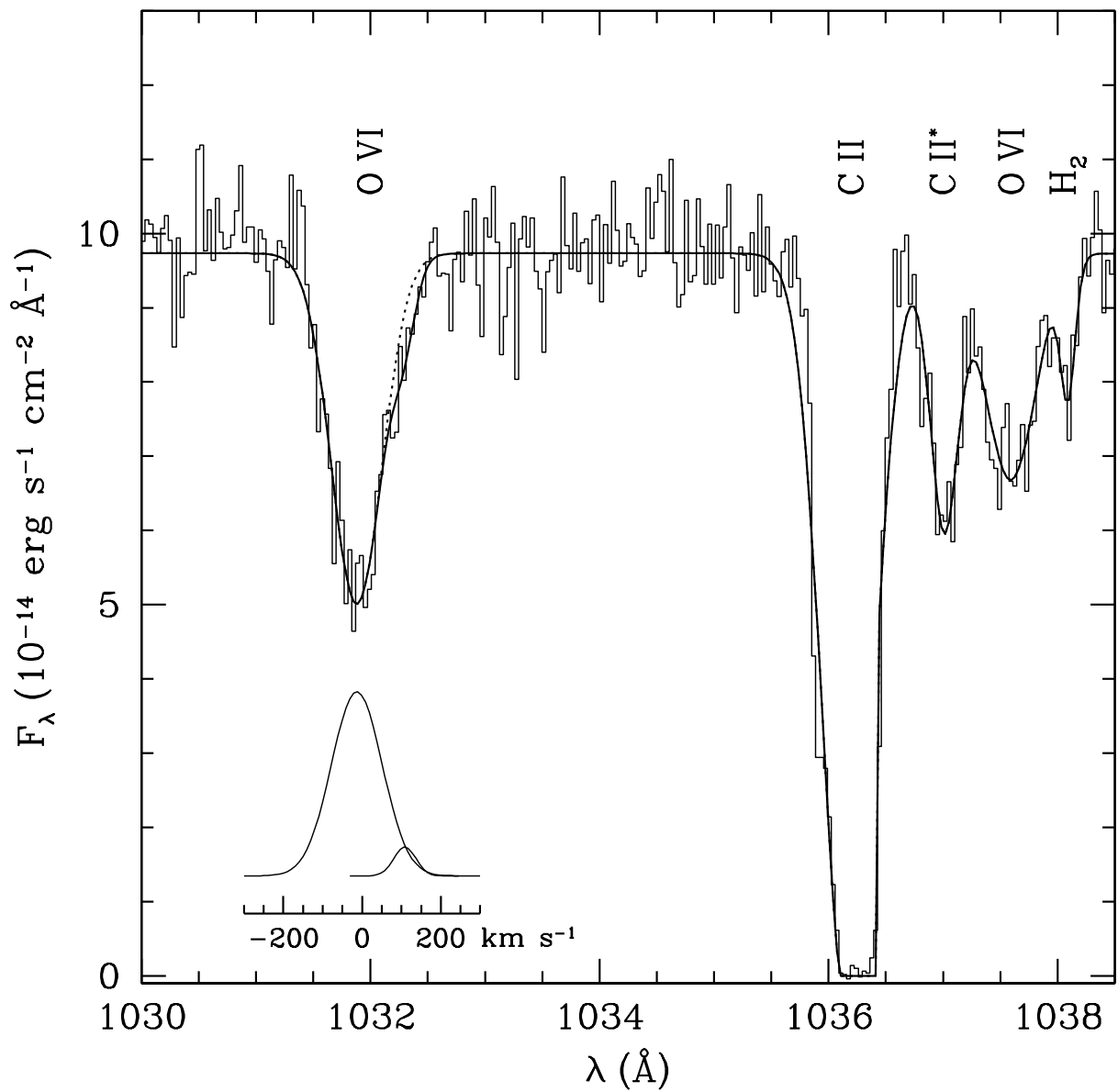}
\caption{\fuse\ spectrum of Mkn~421 around the 
\ion{O}{6}$\lambda\lambda 1032,1037$\,\AA\ absorption doublet (histogram). 
The dark curve shows the 
best--fit model with (solid line) and without (dotted line) the
1032\,\AA\ HVC included.  The inset
plot shows the best--fitting Galactic (large Gaussian) and high--velocity
(small Gaussian) components for the 1032\,\AA\ line, plotted against
velocity relative to the \ovi\ rest wavelength. \label{fig_fuse}}
\end{figure}

The \fuse\ spectrum (Figure~\ref{fig_fuse}) 
shows a strong, broad low--velocity \ovi\ 1032\,\AA\ absorption
line at $z\sim 0$ due to gas in the Galactic thick disk and halo
\citep{savage03}.  An asymmetric wing on the red side of this line is
evident, possibly a kinematically distinct HVC.  We fitted the 
\ovi\ 1032\,\AA\ line in \sherpa\ using a constant local continuum 
(in a $\pm 2$\,\AA\ window) and two Gaussian absorption
components: one for the $v\sim 0$ \ovilv\ line, and one at $v\sim 100\kms$
for the HVC.  No H$_2$ contamination is seen at the \ovi\ 1032\,\AA\ wavelength
when absorption templates are fit to the other H$_2$ lines seen in the
spectrum.  The 1037\,\AA\ line is somewhat blended with a single H$_2$ 
absorption line; this is taken into account with another narrow Gaussian.
From this fit, we find equivalent widths of 
$18.6\pm 5.6$\,m\AA\ for the 1032\,\AA\ HVC and $270.7\pm 7.9$\,m\AA\ for the
Galactic component.  The best--fit model for the HVC is fairly robust
and not sensitive to variations in the initial parameters; however, the
derived equivalent width of $18.5\pm 5.6$\,m\AA\ is lower than the 
$37\pm 11\pm 29$\,m\AA\ (errors are statistical and systematic, respectively)
measured by \citet{wakker03} 
in the initial 21.8\,ks observation.  They employed
a direct integration method which may not have taken into 
account the substantial blending of the Galactic \ovilv\ with the HVC.
Our total \ovi\ equivalent width (LV$+$HVC $=279\pm 10$\,m\AA) 
is in good agreement with their value of $285 \pm 20$\,m\AA.

Deblending the \ovi\ 1037\,\AA\ line is less certain due to the presence
of adjacent Galactic \ion{C}{2} and H$_2$ absorption.  A
flat continuum was again employed from $1035-1040$\,\AA\ and single Gaussian
components were used to fit the \ion{C}{2}, \ovi, and H$_2$ absorption lines. 
The HVC on the 1032\,\AA\ \ovi\ line should also appear at
$\sim 1038$\,\AA\ with $W_\lambda=0.50\times W_\lambda$(1032\,\AA).
Although this component
is too weak to be detected directly, it could cause the measurement of
the Galactic LV--\ovi\ 1037\,\AA\ line to be systematically high.  
Another absorption Gaussian with one--half of the 1032\,\AA\ HVC equivalent
width (and with the same FWHM and velocity offset) was included in the
1037\,\AA\ \ovi\ line fit to account for this.  Table~\ref{tab_lines}
lists the measured properties of the \ovilv\ and HVC absorption lines.

\section{Absorption Line Diagnostics}

\subsection{Doppler Parameters\label{sec_doppler}}
To convert the measured equivalent widths to ionic column densities,
we calculated curves of growth for each absorption line
over a grid of Doppler parameters ($b=10-100\kms$) and column densities
($\log N_H/{\rm cm}^{-2} = 12.0-18.0$), assuming a Voigt line profile.  
Since the X-ray lines are unresolved, $b$ cannot be measured directly.
It can, however, be inferred from the relative strengths of the
three measured \ovii\ K--series lines.  These lines are produced
by the same ionic species, so in an unsaturated medium
$W_\lambda \propto \flu\lambda^2$ where $\flu$ is the oscillator strength.
The expected equivalent width ratio of \ovii\ K$\beta$ to K$\alpha$ is
then $W_\lambda ({\rm K}\beta)/W_\lambda ({\rm K}\alpha)=0.156$, so
the measured value of $0.49\pm 0.09$ indicates that the 
K$\alpha$ line is saturated.  On the other hand, the measured \ovii\ 
${\rm K}\gamma/{\rm K}\beta$ ratio is $0.43\pm 0.16$, in agreement with the
predicted (unsaturated) value of $0.34$.

These line ratios by themselves are insufficient to determine the physical
state of the \ovii--absorbing medium since $b$ and $\novii$ are
degenerate: the K$\alpha$ line saturation could be due to high column
density, low $b$, or a combination of both.  However, given an absorption 
line with a measured equivalent width and known
$\flu\lambda^2$ value, the inferred column density as a function of the
Doppler parameter can be calculated.
The measured equivalent width (and errors) for
each transition thus defines a region in the $\novii-b$ plane.
Since the actual value of $\novii$ is fixed, $b$ and $\novii$ can be
determined by the region over which the contours ``overlap;''  i.e.
the range of Doppler parameters for which the different transitions
provide consistent $\novii$ measurements.

\begin{figure}
\plotone{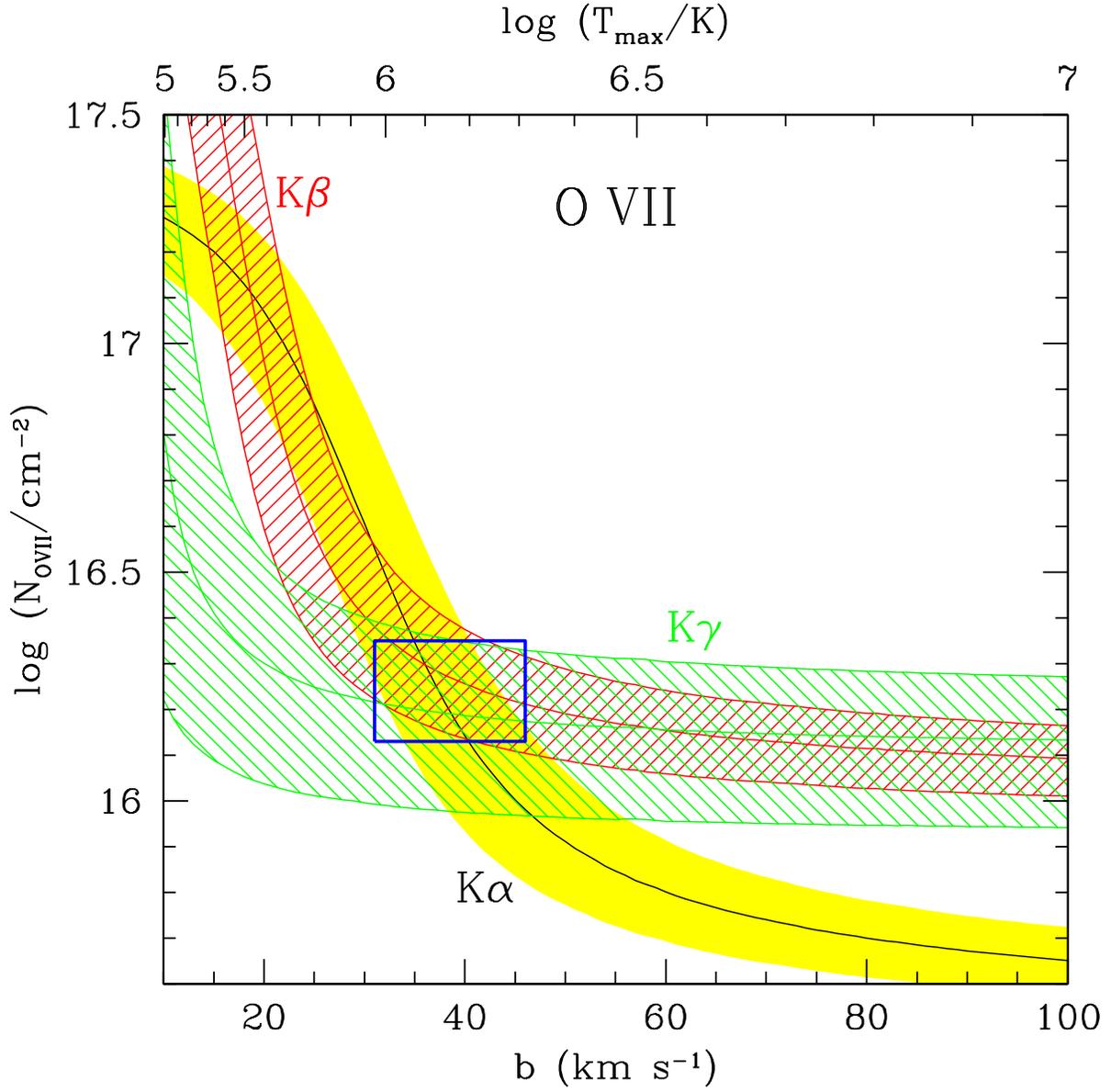}
\caption{Contours of allowed $\novii$ and $b$ for the \ovii\ $K_\alpha$
(yellow), $K_\beta$ (red), and $K_\gamma$ lines.  Shaded regions depict 
the $1\sigma$ errors on $W_\lambda$, with the best-fit $W_\lambda$ line
in the center of each region.  The overlap between the $K_\alpha$
and $K_\gamma$ region provides a $2\sigma$ lower limit of $b>24\kms$, 
and the $K_\alpha-K_\beta$ overlap implies $b<55\kms$.  The weighted
average \ovii\ column density of the three transitions over the $1\sigma$
$b$ range is $\log(\novii)=16.24\pm 0.11$. The blue box depicts the 
$1\sigma$ ranges in $\log(\novii)$ and $b$.  Also labeled on the top axis is
$\log T_{\rm max}=\log (m b^2/2k)$, the temperature inferred from
a given Doppler parameter assuming purely thermal motion. \label{fig_nbovii}}
\end{figure}

Figure~\ref{fig_nbovii} shows such $1\sigma$ contours for the three measured
\ovii\ transitions.  As expected, the inferred $\novii$ is nearly constant
in the unsaturated regime (large $b$), and rises sharply at low $b$ as
the lines begin to saturate.  At each value of $b$, the differences
$\Delta(\log N_{\alpha\beta})= \log(N_{K\alpha})-\log(N_{K\beta})$ and 
$\Delta(\log N_{\alpha\gamma})=\log(N_{K\alpha})-\log(N_{K\gamma})$ 
were calculated, along with the errors on each $\Delta(\log N)$.  The quantity
$\Delta(\log N_{\alpha\beta})$ is consistent with zero at the $1\sigma$
and $2\sigma$ levels for $15<b<46\kms$ and $13<b<55\kms$ respectively, 
while $\Delta(\log N_{\alpha\gamma})$ provides limits of
$31<b<50\kms$ and $24<b<76\kms$ respectively.  Since 
$\Delta(\log N_{\alpha\gamma})$ provides a more stringent lower limit on $b$
while $\Delta(\log N_{\alpha\beta})$ better constrains the upper limit,
we thus assume a $1\sigma$ range of $31<b<46\kms$, and a $2\sigma$ range
of $24<b<55\kms$.  It should be noted that Figure~\ref{fig_nbovii} also
shows some overlap between the $K_\alpha$ and $K_\gamma$ at $b\la12\kms$; 
however, this solution is unlikely given the 
lower limit provided by the $K_\beta$ line.  Moreover, $b=12\kms$ implies
a maximum temperature (assuming purely thermal motion) of 
$T_{\rm max}=1.3\times 10^5$\,K; such a low temperature is unlikely to 
produce the observed strong high--ionization lines.

\begin{figure}
\plotone{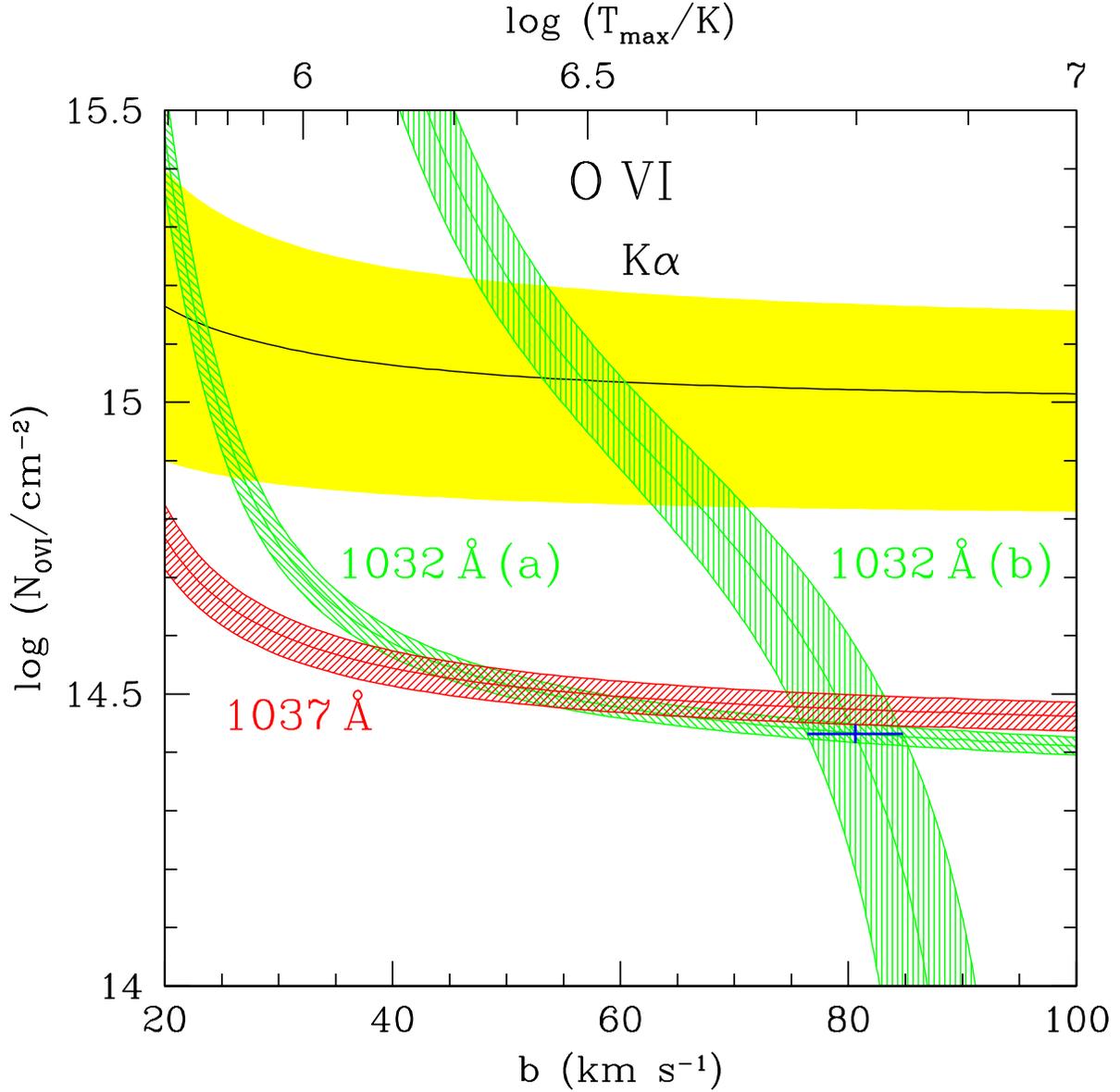}
\caption{Contours of allowed $\novi$ and $b$ for the \ovi\ 1032\,\AA\ (green),
1037\,\AA\ (red), and putative K$\alpha$ 22\,\AA\ (yellow) lines.  Contour (a)
is derived from the 1032\,\AA\ \ovilv\ equivalent width while (b) is from
the measured FWHM; the intersection between the two green contours provides
a tight constraint of $b=80.6\pm 4.2\kms$ and 
$\log \novi({\rm cm}^{-2})=14.432\pm 0.016$ for the \ovilv\ (shown as the
blue cross).  Note that the \ovi\ K$\alpha$ 
transition predicts $\novi$ about 0.5 dex higher than the UV line, and 
this discrepancy cannot be explained by saturation. \label{fig_nbovi}}
\end{figure}

A similar analysis is not as effective when applied to the strong \ovilv\ 
UV doublet (from the thick disk),
since these lines are only slightly saturated: the measured $W_\lambda$
ratio is $0.61\pm 0.04$, compared to the expected unsaturated value of 0.50.
When the inferred $\novi$ is calculated as a function of $b$ for both lines
of the \ovilv\ doublet, the predicted $\novi$
values are consistent over $b= 34-112\kms$ (at the $2\sigma$ level; see 
Figure~\ref{fig_nbovi}).  Since the \ovilv\ 1032\,\AA\ line is fully resolved 
by \fuse\ ($\sim 15$ resolution elements across the line profile)
and relatively unblended, its Doppler parameter can be estimated much more
accurately using the measured line width and strength.  In an unsaturated
absorption line, ${\rm FWHM}=2(\ln 2)^{1/2}b$; however, the measured FWHM
increases if the line is saturated.  We compensated for this by 
calculating Voigt profile FWHMs on a grid of $\novi$ and $b$,
and determining the region consistent with the \ovilv\ 1032\,\AA\ FWHM 
measurement of $152\pm 7\kms$ (or $b=91\pm 4\kms$ assuming no saturation).

When the FWHM--derived contour is overlaid on the $\novi-b$ contour inferred
from the equivalent width measurement of the LV--\ovi\ 1032\,\AA\ line, 
the two regions overlap nearly orthogonally 
(Figure~\ref{fig_nbovi}) leading to a constraint of 
$b$(\ovilv)$=80.6\pm 4.2\kms$.  This is $\sim 2\sigma$ lower than the 
unsaturated FWHM, once again confirming that the \ovilv\ is only
weakly saturated.  At this $b$ the inferred column densities
from the two lines of the \ovilv\ doublet differ by $1.3\sigma$ 
but this is only a minor discrepancy and likely due to errors introduced
by the blending of the 1037\,\AA\ line; thus, we will only consider results
from the more reliable 1032\,\AA\ line measurement.  However, at no value
of the Doppler parameter do the 1032\,\AA, 1037\,\AA, and \ovi$_{K\alpha}$
lines all produce a consistent $\novi$ measurement; in fact, the
\ovi$_{K\alpha}$ column density is a factor of $\sim 4$ higher than that
inferred from the UV data.  This discrepancy is discussed further
in \S\ref{sec_ovi}.

\subsection{Column Densities}

The Doppler parameters measured for the \ovii\ ($31\kms < b < 46\kms$) 
and \ovilv\ ($b=80.6\pm 4.2\kms$) absorption
are inconsistent at the $\sim 3\sigma$ level, indicating the presence
of at least two distinct components: the Galactic thick--disk
gas traced by broad $v\sim 0$ \ovilv\ absorption, and another
lower--$b$ phase, possibly of extragalactic origin, traced by the
\ovii\ absorption lines.  It cannot be assumed {\it a priori} 
that any given line (other than those used to determine $b$)
originates in one phase or another; moreover, 
the uncertainty in the calculated column density
depends not only on the equivalent width error but also the error in $b$.
To take this into account, for each ion the derived column density 
$\log N_i$ and 
its $\pm 1\sigma$ limits were averaged over the $\pm 1\sigma$ 
ranges of both measured Doppler parameters. As it turns out, the choice 
of $b$ does not make a significant difference since all other lines 
(besides the \ovii\ and \ovilv\ absorption) are essentially unsaturated;
i.e., the difference in $N_i$ calculated with the \ovilv\ and \ovii\ Doppler
parameters is 
small compared to the $1\sigma$ error on the equivalent width measurements. 
Even so, to avoid possible systematic errors, we assumed $b= 80.6\pm 4.2\kms$
for those lines {\it likely} to originate in the Galactic thick disk 
(\ovilv, \ov, 
and \cv), and $b=31-46\kms$ for all other species.  The derived ionic
column densities are listed in Table~\ref{tab_lines}

\subsection{Temperature and Density Constraints}

At densities such as those found in the Galactic interstellar medium (ISM;
$n_e \sim 1$\,cm$^{-3}$), photoionization is unimportant because 
thermal collisions are by far the dominant ionization source.
This is also the case for very high temperatures ($T\ga 10^7$\,K) even at
low densities, since
the collisional rate is greater than the photoionization rate.
However, at the low densities typically found in the intergalactic medium 
($n_e = 10^{-6}-10^{-4}$\,cm$^{-3}$), 
photoionization from the diffuse UV/X-ray background begins to play a greater
role by enriching the abundances of highly--ionized elements
at typical WHIM temperatures ($\log T({\rm K})\sim 5-7$) relative to 
those expected from pure collisional ionization \citep{nicastro02,mathur03}.  
It is thus imperative that the ionizing background be taken into account 
in order to accurately predict ionic abundances in the WHIM.

Version 90.04 of the ionization balance code {\it Cloudy} 
\citep{ferland96} was used to compute collisional-- plus photoionization
hybrid models for the absorbing medium.   Relative ionic abundances 
were computed over a grid of $\log T({\rm K})=4.5-7.4 $ and 
$\log n_e({\rm cm}^{-3})=-7-0$, with a step size of 0.1~dex in both
$\log n_e$ and $\log T$.  Initially,
a rigid scaling of $[{\rm Z/H}]=-1$ for all metals was assumed.
For the ionizing background we employed the 
\citet{sternberg02} fit to the metagalactic radiation field:
\begin{equation}
J_\nu = \left\{\begin{array}{ll}
J_{\nu 0}\left(\frac{\nu}{\nu_0}\right)^{-3.13} & 1<\frac{\nu}{\nu_0} < 4\\
2.512\times 10^{-2} J_{\nu 0} \left(\frac{\nu}{\nu_0}\right)^{-0.46} & 
\frac{\nu}{\nu_0} > 4\\
\end{array}\right.
\end{equation}
where here 
$J_{\nu 0} = 2\times 10^{-23}$\,ergs\,s$^{-1}$\,cm$^{-2}$\,Hz$^{-1}$\,sr$^{-1}$
and $\nu_0=13.6$\,eV.  The total flux of ionizing photons is then 
given by
$f_\gamma = 4\pi \int_{\nu_0}^\infty (J_\nu/h\nu) d\nu
= 1.3\times 10^4$\,photons\,s$^{-1}$\,cm$^{-2}$, and 
the ionization parameter is 
$\log U= \log (f_\gamma/c) - \log n_e=-6.36-\log(n_e)$ where $n_e$ is
the electron density in cm$^{-3}$.

Using the ionic abundances calculated with {\it Cloudy}, we derived
expected abundance ratios
for all observed ions at each point in the $\log n_e-\log T$ plane.  
Since any given density and temperature uniquely determines a set of abundance
ratios ($N_a/N_b$ for all ions $a$ and $b$), the problem can be inverted: 
any value of $N_a/N_b$ defines a curve in the $\log n_e-\log T$ plane, 
i.e. a set of temperatures and densities which can produce the measured
ratio.  When the errors on $N_a/N_b$ are taken into account, the curves
become contours, and the overlap between two or more contours defines
the temperatures and densities for which the measured ratios are consistent.
This is analogous to the method used in \S\ref{sec_doppler} to determine
Doppler parameters for \ovilv\ and \ovii.

\begin{figure}
\plotone{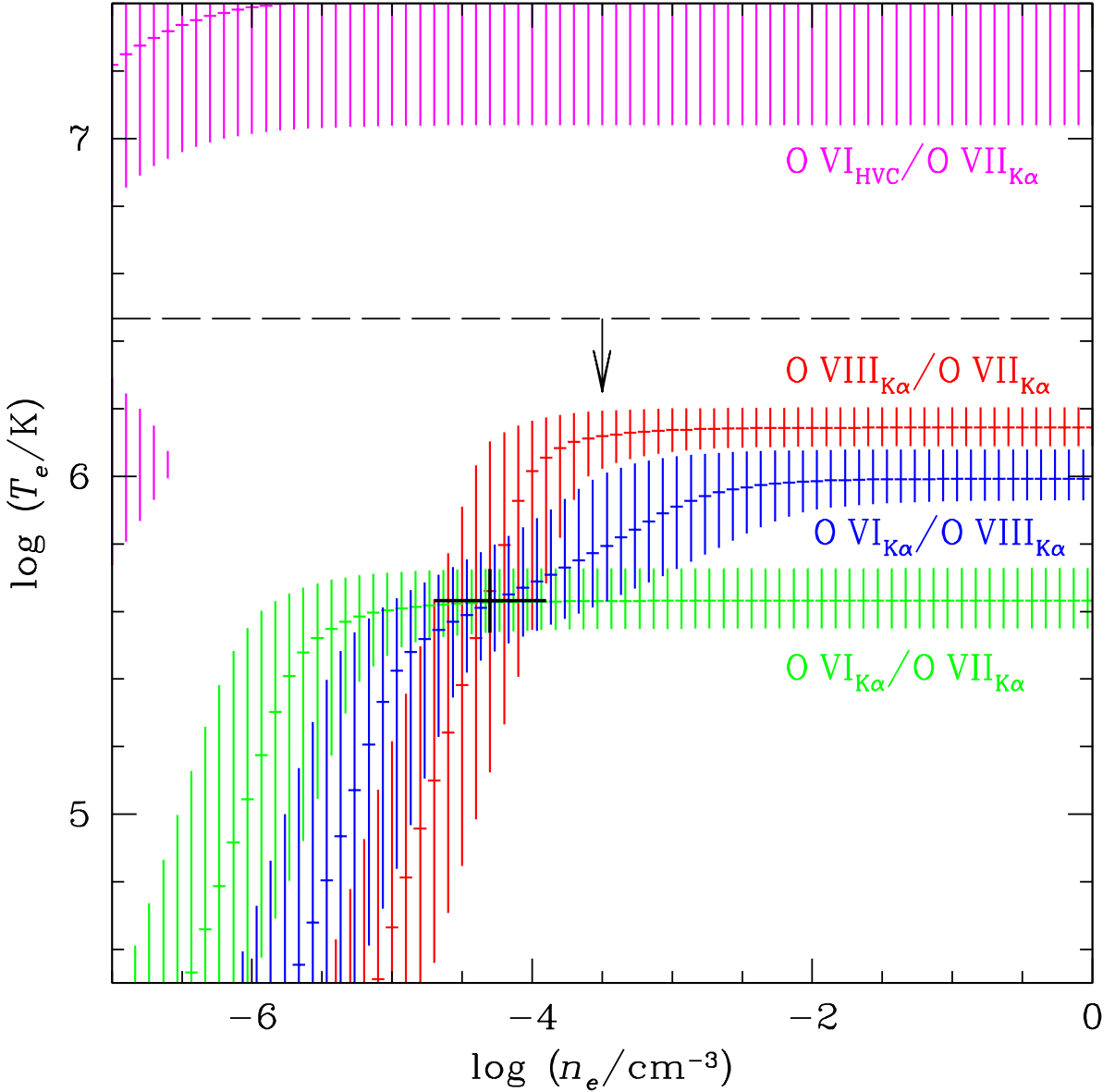}
\caption{Contours of constant abundance ratios for X-ray and UV oxygen
absorption lines.  Vertical bars denote the $2\sigma$ range of temperatures
inferred from the abundance ratio at each step in $\log n_e$.  The horizontal
dashed line is the $2\sigma$ upper limit on the temperature of the \ovii\
absorber from the Doppler parameter measurement.  While the X-ray line ratios
predict significantly different temperatures in the collisional ionization 
regime ($\log n_e \ga -3$), they are consistent with 
$\log n_e (\rm{cm}^{-3})=-4.7$ to $-3.9$ and $\log T(K)\sim 5.5-5.7$ (denoted
by the black cross).
Note that the \ovi$_{\rm HVC}$/\ovii\ ratio predicts a temperature far
higher than the other derived limits, and thus the HVC is unlikely to
be associated with the \ovii\ absorption. \label{fig_oxplot}}
\end{figure}

The most powerful diagnostics are those using ratios between different 
ions of the same element, since these ratios are independent of the relative
metal abundances.
Unfortunately the \nvii/\nvi\ and \nex/\neix\ upper limits
are not stringent enough to place meaningful constraints on the 
temperature and density.
Figure~\ref{fig_oxplot} shows the $\log n_e-\log T$ contours for ratios
between the X-ray \ovi$_{K\alpha}$, \ovii, and \oviii\ lines as well as the 
\ovi$_{\rm HVC}$/\ovii\ ratio.
The X-ray line ratios are inconsistent with a high--density
($n_e\ga 10^{-3}$cm$^{-3}$), high--temperature ($\log T>6.2$)
medium, and instead converge on a partially
photoionized plasma with $n_e=10^{-4.7}-10^{-3.9}$\,cm$^{-3}$ (from
the overlap between the \ovi$_{K\alpha}$/\ovii\ and \oviii/\ovii\ contours)
and $T=10^{5.5-5.7}$\,K (from the limits provided by \ovi$_{K\alpha}$/\ovii\ 
in this density range).  These ranges of temperatures and densities are
in line with those expected from WHIM gas \citep{dave01}.  
Of course, this is all 
contingent on the \ovi$_{K\alpha}$ line being a reliable tracer of 
$\novi$; this caveat is discussed in detail in \S\ref{sec_analysis}.

On the other hand, the \ovi$_{\rm HVC}$/\ovii\ 
ratio overpredicts the temperature by at least an order of magnitude for
all values of $\log n_e$---in order to be consistent with the \oviii/\ovii\
ratio, the \ovi$_{\rm HVC}$/\ovii\ ratio would need to be stronger 
by a factor of $\sim 2.5$ (or 
$\sim 3.5\sigma$).  It is possible that the HVC is not a physically distinct
component, but is instead the result of some systematic error (such as
fixed pattern noise or an unexpected anomaly in the Galactic \ovilv\ 
velocity distribution).
In this case, the \ovi\ associated with the \ovii\ and \oviii\ may
be completely blended with the thick--disk \ovilv\ and thus unmeasurable.
Consistency with the \oviii/\ovii\ ratio (in the collisional ionization
regime) requires $\log(\novi/\novii)\sim -2.5$,
or roughly 20\% of the Galactic UV \ovi\ absorption.
Although it appears that the \ovi\ HVC as measured cannot originate
in the same medium as the \ovii\ absorption, we suspect that additional
atomic physics may be at work here and could in principle reconcile 
this disagreement (see \S\ref{sec_ovi}).

While the \ovi$_{K\alpha}$/\ovii\ and \oviii/\ovii\ ratios provide strong
constraints, it is important to consider other ion ratios as well
(particularly since the \ovi$_{K\alpha}$ and 1032\,\AA\ \ovi\ column
densities disagree).  Figure~\ref{fig_allplot} shows the $\log n_e-\log T$
contours for several different ion ratios, all calculated relative to \ovii\
since the error on $\novii$ is small.  With a rigid metallicity shift
relative to Solar, the \neix/\ovii\, \oviii/\ovii, and 
\ovi$_{K\alpha}$/\ovii\ ratios are not all consistent with each other for
any combination of temperature and density; however, the consistency can
be improved with adjustments to the [Ne/O] ratio (see \S\ref{sec_abundances}).
Both the \cvi/\ovii\ and \nvii/\ovii\ measurements are consistent with a
high-- or low--density medium at solar abundances.

Limits on the temperature of the Galactic thick--disk absorption can be 
derived in a 
similar fashion, although it is not the primary focus of this work and 
there are far fewer
measured lines to work with.  The most accurately--measured line is
the \ovilv; additionally, \cv\ and \nvi\ X-ray lines are measured, and
upper limits have been determined for \ov\ and \nvii. 
Figure~\ref{fig_galewtplot} shows the temperature constraints derived
for this Galactic absorption, assuming pure collisional ionization.
The \ov/\ovilv\ and \nvii/\nvi\ upper limits provide 
metallicity--independent constraints of $\log T > 5.39$ and
$\log T < 6.64$ respectively.  A more stringent upper limit on
temperature of $\log T < 6.03$ is provided by the \nvii/\ovilv\ ratio,
but this is somewhat dependent on [N/O].  Within this range the \cv/\ovilv\
ratio provides an even stricter limit of $5.3<\log T<5.7$, but again
this depends on [C/O].  At these temperatures the expected 
\ovii\ column density is at most an order of magnitude less than measured;
thus, the \ovilv, \ovii, and \oviii\ cannot all originate in the same
phase assuming pure collisional ionization \citep[see also][]{mathur03}.

\section{Discussion\label{sec_analysis}}
Our \chandra\ and \fuse\ observations have provided a wealth
of data on absorption near the Galaxy, constraining the temperature and
density tightly ($\log T({\rm K})=5.5-5.7$ and 
$n_e=10^{-4.7}-10^{-3.9}$\,cm$^{-3}$ when the \ovi$_{K\alpha}$ measurement
is included), which are conditions suggestive of the local group intergalactic
medium and require supersolar [Ne/O]. Here we first examine the assumptions
that have led us to these results (\S\ref{sec_caveats}), 
and then we discuss their implications
for the location of the absorbing gas (\S\ref{sec_origin}), 
subject to these caveats.

\subsection{Potential Caveats\label{sec_caveats}}
\subsubsection{The \ovi\ Discrepancy\label{sec_ovi}}
The interpretation of the UV and X-ray data are particularly important,
since (as Figure~\ref{fig_oxplot} shows) the combined \oviii/\ovii\ and
\ovi/\ovii\ ratios can provide tight constraints on the absorber
temperature and density simultaneously 
\citep[see also Figure~5 in][]{mathur03}.  However, in this case the
\ovi\ column density inferred from the \ovi$_{K\alpha}$ is a factor
of $\sim 4$ higher than the combined 1032\,\AA\ low-- and high--velocity
components.  Since both the X-ray and UV transitions trace the
same atomic state, the inferred column densities should match.  A similar
disagreement has been seen in intrinsic AGN absorption systems
\citep[see][]{krongold03,arav03}; however, in these cases it is typically
attributed to saturation or a velocity--dependent covering factor, neither
of which is relevant to this $z\sim 0$ absorption.  

On the other hand, our \ovi$_{K\alpha}$ measurement provides a test
for these attributions; the local absorption, after all, is likely a 
dramatically different physical system than an AGN outflow, yet the same
conflict arises.  A macroscopic explanation does not adequately describe 
how this discrepancy 
is seen in both physical systems, so the actual reason may lie in
the atomic physics of highly ionized plasmas.  For example, some fraction
of the \ovi\ 
may be excited through collisions or recombination from \ovii,
and thus unable to produce 1032\,\AA\ absorption while still absorbing
\ovi$_{K\alpha}$ photons.  While a scenario that produces significant
depopulation of the \ovi\ ground state is difficult to envisage in such
a low density plasma, we are investigating further the statistical
equilibrium of \ovi\ including photoexcitation and recombination in order
to study such effects in more detail.  
However, it should be emphasized that this is not
an isolated case so there must be a physical explanation for the \ovi\ 
discrepancy, and the resolution of this paradox is crucial to our 
understanding of \ovi\ UV and X-ray absorption and how it relates to
the \ovii.

There is also the possibility that the line was misidentified
as \ovi$_{K\alpha}$, and is actually another intervening \ovii\ 
absorption line at $z=0.0195$.  This
latter explanation is unlikely since no other absorption lines at this
redshift are seen in the \fuse\ or \chandra\ spectra; additionally,
this would require the line to fall exactly on the
\ovi\ rest wavelength, which seems like an improbable coincidence.  
Another possibility is that the theoretical oscillator strength of the
\ovi$_{K\alpha}$ transition is incorrect, but the value given in
\citet{pradhan00} would need to be
low by a factor of $\sim 2-4$, in sharp contrast to the successful calculations
of $f_{lu}$ for inner shell transitions in other ions in the same paper.  
Nevertheless, due to the discrepancy between the UV and X-ray \ovi\ column
density measurements, we
present both possibilities: either (a) the \ovi$_{K\alpha}$ line
measures $\novi$, or (b) it does not and is thus ignored.

\subsubsection{Absorption Components}
The Doppler parameter measurements indicate the existence of two distinct
components along the Mkn~421 line of sight: one seen in the thick--disk
\ovilv\ 1032\,\AA\ absorption with $b_{\rm LV}=80.6\pm 4.2\kms$, and the \ovii\ 
absorber
with $b_{\rm OVII}=31-46\kms$ ($1\sigma$ limits).  The \ovi\ HVC 
may represent a third phase (if case (b) above is correct)
with $b_{\rm HVC}=35^{+18}_{-10}\kms$ (from
the FWHM measurement).  This agrees surprisingly well with the \ovii\
$b$ measurement, and is consistent with numerical simulations of
the nearby IGM \citep{kravtsov02}.  However, the extremely low
\ovi$_{\rm HVC}$/\ovii\ ratio requires a temperature much higher than
the upper limit provided by $b_{\rm OVII}$.  In order for the
HVC to trace the same gas as \ovii\ (case a), then, the aforementioned
atomic physics effects would need to be suppressing \ovi\ HVC absorption
and not the \ovilv\ line.
\citet{sembachetal03} list mean Doppler parameters for a variety
of HVCs, both Galactic and probable Local Group; unfortunately, the 
dispersion in these values and the errors
on $b_{\rm OVII}$ and $b_{\rm HVC}$ measured here are both too large 
to associate the components presented here to one of their classifications.

It is also important to note that our analysis assumes a single phase
origin for the included X-ray lines.  This assumption is consistent with
the data, given the good agreement between the three \ovii\ lines in
the calculated ranges of $b$ and $\novi$\ (Figure~\ref{fig_nbovii}).
Even so, if any of the ionic species arises
in more than one phase along the line of sight, our results could be affected.
For example, a Galactic, purely collisionally ionized medium can
in principle reproduce the observed relative abundances of \oviii, \ovii,
\ovi$_{K\alpha}$, and \neix\ if several unresolved components are invoked 
to explain this
absorption.  However, the simplest explanation (a single--phase, low--density,
partially photoionized extragalactic absorber) is fully
consistent with all of these line measurements, and the similarity of the 
derived absorber properties to expectations for the local WHIM lend
further support to this model (see \S\ref{sec_origin}).

\subsubsection{Abundances\label{sec_abundances}}
Due to the uncertainty associated with the \ovi$_{K\alpha}$ absorption,
metal abundances relative to oxygen play a particularly important role in
this analysis.
By adjusting the metal abundances of the 
absorbing gas, the consistency of the solutions can in principle be improved
with and without the \ovi$_{K\alpha}$ measurement.
Although the $\log n_e-\log T$ contour plots are useful for finding
solutions, they cannot easily be used to determine the effects of changes
in elemental abundances; thus, Figure~\ref{fig_ewtplots} shows $N_i/\novii$ as
a function of $\log T$ for both the $\log n_e=-3.9$ and $\log n_e=0$
cases.  In this figure, the $y$--ranges given by the measured ratios 
(thick lines) shift up and down as a result of decreases and increases
in the abundances relative to oxygen, respectively; thus different parts
of the theoretical curves would be in bold, moving the allowed temperature
ranges (shown in the lower panel) to the left or right.  Solar abundances
here are taken to be the {\it Cloudy}~90 defaults \citep{ferland96,grevesse89}.

\begin{figure}
\plottwo{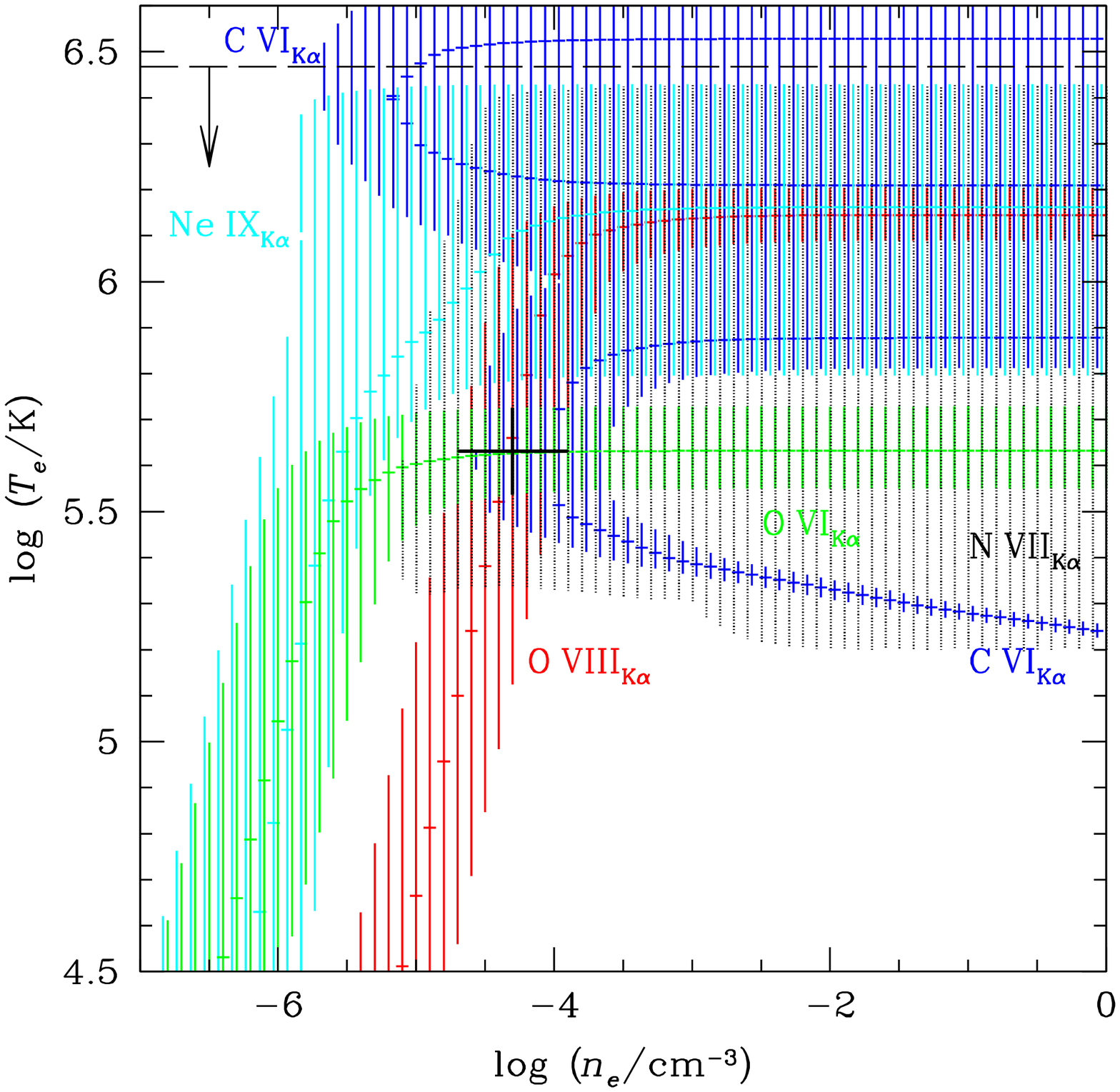}{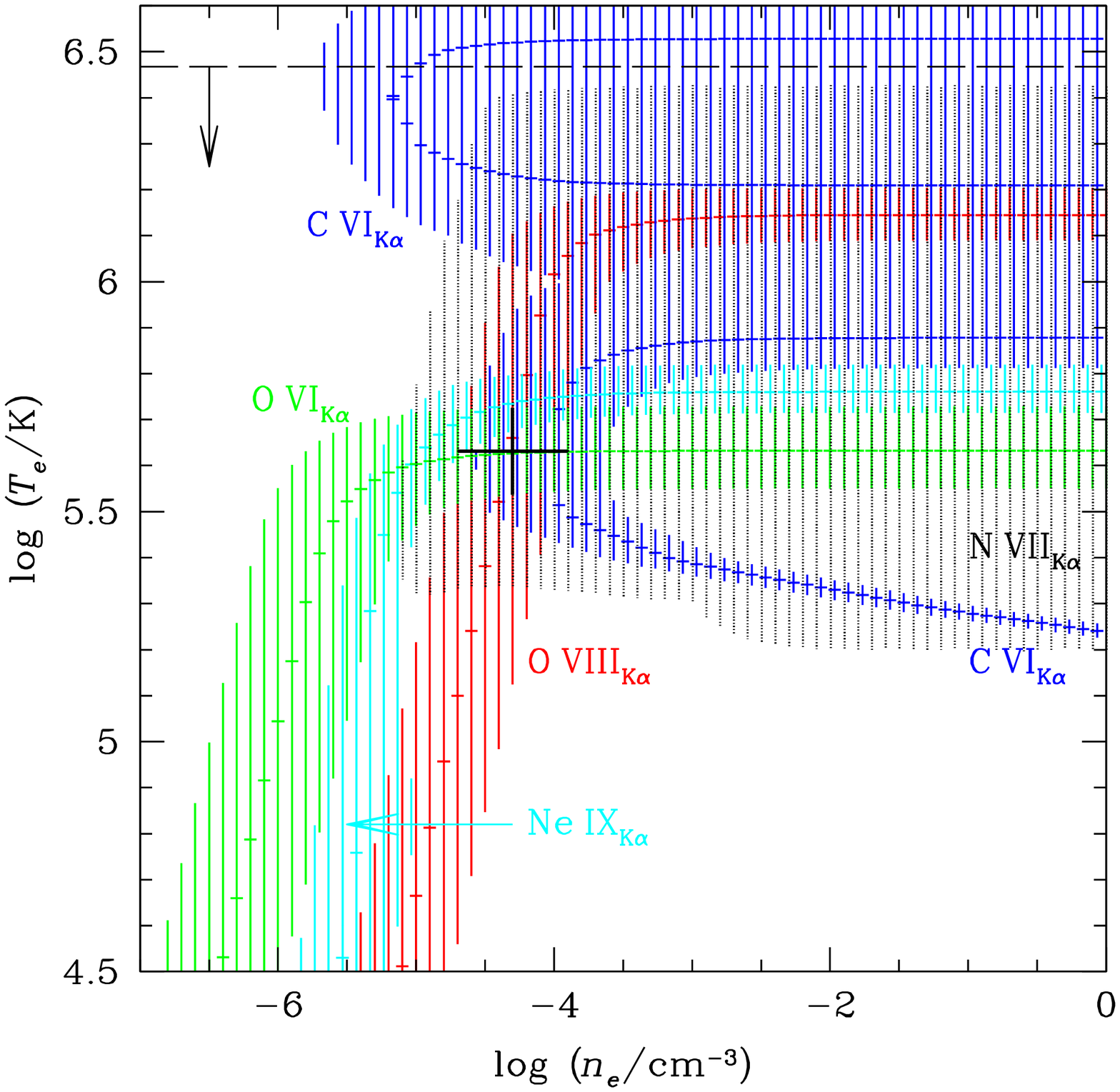}
\caption{Same as Figure~\ref{fig_oxplot}, but for ratios of several different
ion abundances to \ovii: \ovi$_{K\alpha}$ (green), \oviii\ (red), 
\cvi\ (blue), \neix\ (cyan), and \nvii\ (dotted black region).  The black
cross shows the range of $\log T$ and $\log n_e$ derived from the 
\oviii/\ovii\ and \ovi$_{K\alpha}$/\ovii\ ratios.  Assuming solar abundances
(left panel), the \ovi$_{K\alpha}$ and \neix\ contours are inconsistent
for $\log n_e\ga -5$, while all other ratios (except \ovi$_{K\alpha}$\ovii)
are consistent at $\log n_e\ga -4.5$.  The right panel
shows how a neon abundance shift of [Ne/O]$=1$ produces better agreement
in the low--density regime.  High--density models agree with the data only 
if the \ovi$_{K\alpha}$ measurement is disregarded.  \label{fig_allplot}}
\end{figure}
\begin{figure}
\plotone{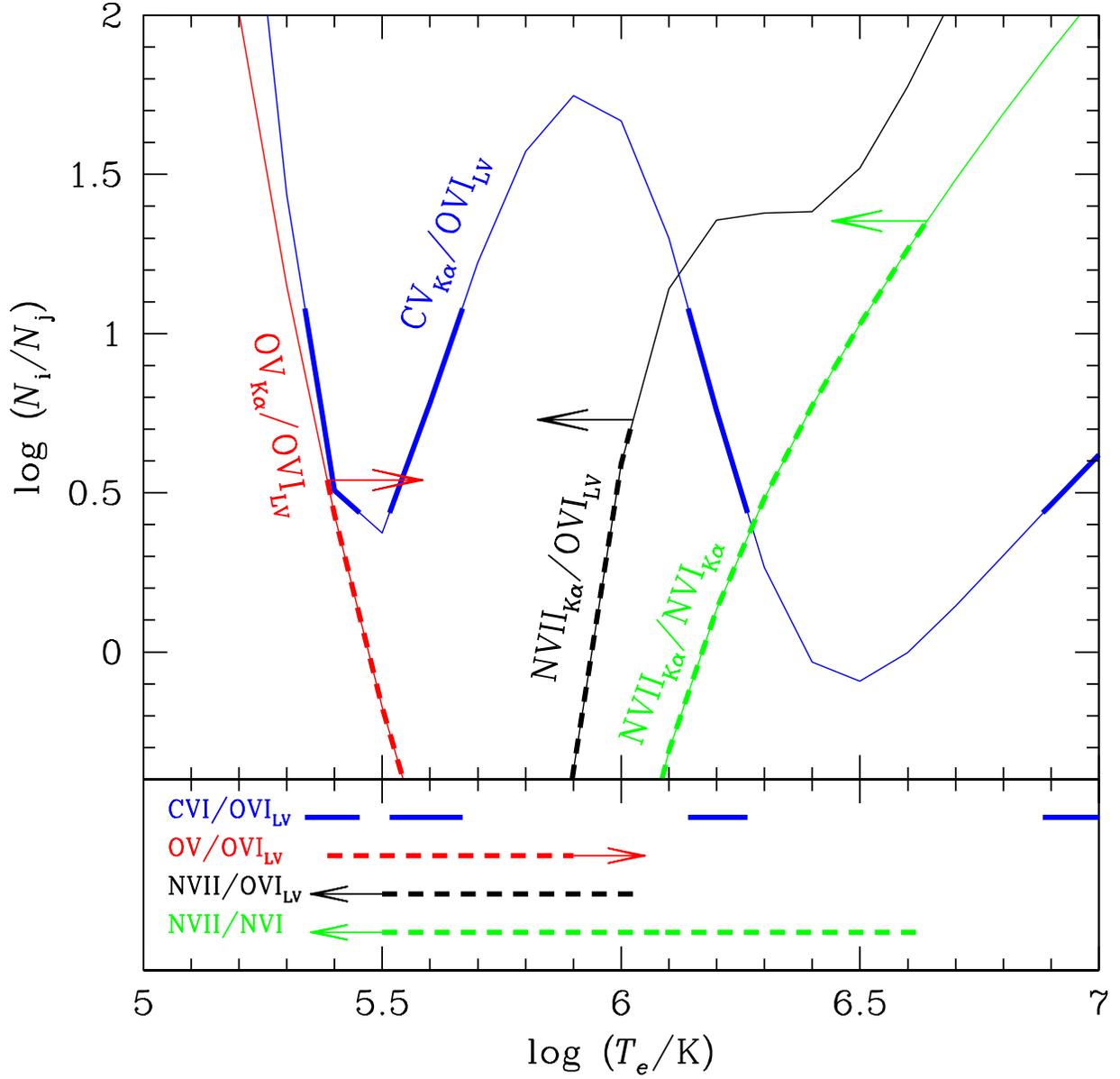}
\caption{Models of ionic column density ratios for ions likely
to arise in the Galactic ISM (assuming pure collisional ionization). Calculated
ratios are shown as thin lines with $\pm 2\sigma$ ratio measurements
overplotted (thick segments).  Upper limits are shown as dashed lines, and
the temperature ranges derived from the different ion ratios are shown in the
lower panel.  The same--element ratios in this case provide strong 
constraints of $\log T>5.4$ (\ov/\ovi) and $\log T<6.6$ (\nvii/\nvi).
\label{fig_galewtplot}}
\end{figure}
\begin{figure}
\plottwo{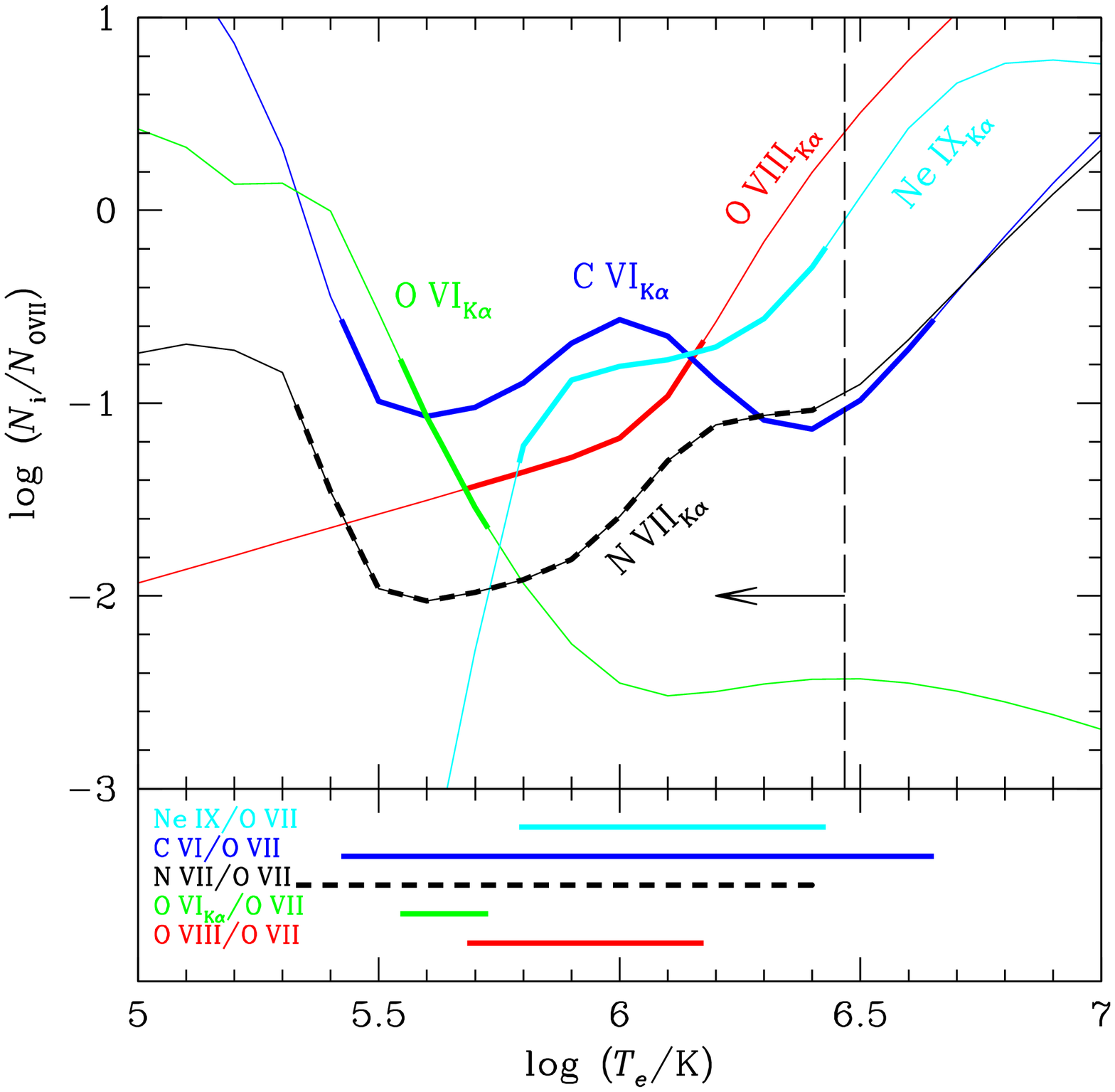}{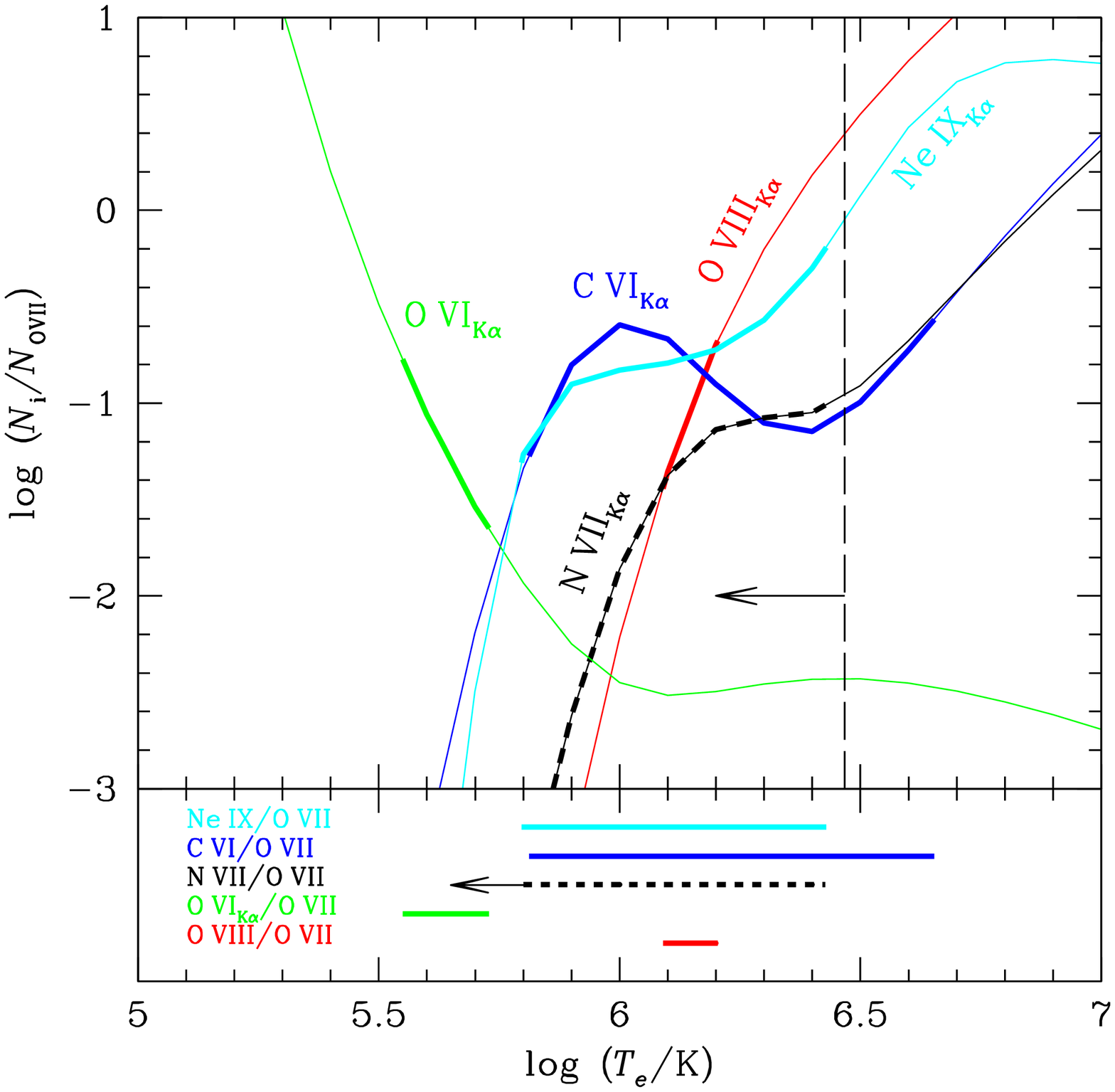}
\caption{Same as Figure~\ref{fig_galewtplot}, for the ions likely originating
in an extragalactic medium, with $\log n_e=-3.9$ (left)
and $\log n_e=0$ (right).  Solar abundances relative to oxygen are assumed,  
and all calculated ratios are relative to $\novii$.
The low--density
case produces a consistent temperature solution of $T\sim 10^{5.7}$\,K 
(from the overlap between the \oviii\ and \ovi$_{K\alpha}$ bold regions), and
[Ne/O] must be supersolar.  The high--density case, on the other hand,
provides a consistent solution with Solar abundances but the temperatures
implied by the \oviii/\ovii\ and \ovi$_{K\alpha}$/\ovii\ ratios are 
inconsistent.
\label{fig_ewtplots}}
\end{figure}

In case (a), i.e.~if the \ovi$_{K\alpha}$ measurement of $\novi$ is correct, 
then the temperature and density of the absorber are tightly
constrained in a metallicity--independent manner, and relative abundances
for other elements can be estimated.  Under this assumption, the
oxygen ion ratios are consistent within a range of 
$n_e=10^{-4.7}-10^{-3.9}$\,cm$^{-3}$; however at solar abundances the 
\neix/\ovii\ ratio demands higher temperatures than allowed by the
\ovi$_{K\alpha}$/\ovii\ ratio.
Over this range of densities, the permitted abundances of
neon, carbon, and nitrogen abundances relative to oxygen (that is, 
the range of abundances which produce line ratios consistent with both
\ovi$_{K\alpha}$/\ovii\ \emph{and} \oviii/\ovii) are then
$0.6\le$[Ne/O]$\le 2.2$, $-0.8\le$ [C/O]$\le 0.3$, and [N/O]$\le 0.9$.  Note
that supersolar [Ne/O] has also been observed in the $z=0$ absorber
toward PKS2155--304 \citep{nicastro02}.
The improvement in the fit from supersolar [Ne/O] is shown in the right panel
of Fig~\ref{fig_allplot}.  Note, however, that in this case the 
discrepancy between the \ovi$_{K\alpha}$ and 1032\,\AA\ measurements
becomes even more severe.  Since the bulk of the \ovilv\ cannot be
associated with the \ovii\ due to the different Doppler parameters,
the UV \ovi\ component associated with the WHIM (hence the \ovi$_{K\alpha}$)
must be substantially 
weaker than the Galactic \ovi\ absorption; thus the discrepancy is
correspondingly larger.

On the other hand, if (b) the \ovi$_{K\alpha}$
line is ignored then the relative abundances in the absorber can be adjusted
such that the measured line ratios are consistent with either a low-- or 
high--density absorber.  As shown in Figure~\ref{fig_allplot}, a high--density,
collisionally--ionized medium fits the data with Solar abundances.  Assuming
this is the case, the temperature is then completely constrained by the
\oviii/\ovii\ ratio at $\log T=10^{6.1-6.2}$\,K, and the relative abundances
consistent with the \oviii/\ovii\ ratio in this temperature range are
$-0.6\le$[Ne/O]$\le 0.6$, $-0.6\le$[C/O]$\le 0.3$, and [N/O]$\le 0.4$.

The requirement
that [Ne/O] be supersolar does not affect the viability of the 
partially--photoionized
model: in both the ISM and low--$z$ IGM, [Ne/O] is observed to be 
significantly larger than the solar value 
\citep[e.g.][]{paerels01,nicastro05a}.  This may be an intrinsic property
of the enriched gas ejected into the IGM and ISM, or instead could be 
due to depletion
of C, N, and O onto dust grains in supernova ejecta
or quasar winds \citep{whittet92,elvis02}.  If the dust destruction timescale
is long and the IGM is continuously enriched by this latter mechanism, 
then the observed supersolar [Ne/O] would be expected.  All of these 
enrichment scenarios are quite uncertain, but few (if any) are able to
produce [Ne/O]$<0$.  Indeed, the solar neon
abundance itself is quite uncertain since it is inferred from solar
wind measurements.  The increase in the solar neon abundance proposed
by \citet{drake05} is supported by these observations, and may
provide another physical argument for the lack of subsolar [Ne/O].

\subsection{Where does the X-ray absorption originate?\label{sec_origin}}
Assuming the \ovii\ absorption system is homogeneous, its radial
extent can be estimated by calculating $r\approx N_H/(\mu n_e)$, where
$\mu\sim 0.8$, $\log(n_{\rm O}/n_{\rm H})=-3.13$ (solar abundance), and 
\begin{equation}
N_H=N_{\rm OVII}\times \left(\frac{N_{O,tot}}{N_{\rm OVII}}\right)
\times 10^{3.13-[{\rm O/H}]}.
\end{equation}
The second term in the equation is approximately unity, since over the
range of temperatures and densities implied from the \oviii/\ovii\ 
ratio, \ovii\ is the dominant ionization state by at least an order
of magnitude; thus, 
$\log N_H=\log \novii+3.13-[{\rm O/H}]=\{20.37-([{\rm O/H}]+1)\}\pm 0.11$.
The measurement error on $\novii$ is small compared to the uncertainty
range in $n_e$, so it can be disregarded.  Assuming that (a) the
\ovi$_{K\alpha}$ line does measure $\novi$, the $2\sigma$ range
of densities is $-4.7\le \log n_e \le -3.9$,
resulting in a radial extent of $r=(0.8-4.9)\times 10^{-([{\rm O/H}]+1)}$\,Mpc. 
These radial extents are consistent with those expected from a Local Group 
medium or local filament interpretation for this absorption 
\citep{nicastro02,nicastro03} and too large to be confined within a Galactic
halo.  The absorber extent could be marginally consistent with a Galactic
corona if the metallicity is high ($r=80-490$\,kpc, $2\sigma$ limits at 
Solar metallicity); however, such a scenario seems implausible, particularly
if this corona primarily consists of gas accreted from the metal--poor IGM.

It is unlikely that this absorption system, if extragalactic, is spherically
symmetric (particularly in a ``local filament'' interpretation). The
total mass in the \ovii\ system can be written as
$M_{\rm tot}=f\times (4/3)\pi r^3 (1.4 n_{\rm H} m_{\rm H})$, 
where $f$ ($<1$) parameterizes the departure from spherical symmetry and 
$1.4n_{\rm H} m_{\rm H}\approx n_{\rm H} m_{\rm H}+n_{\rm He} m_{\rm He}$.  
Replacing $r$ with $N_H/(\mu n_e)$ and plugging in relevant values, 
\begin{equation}
M_{\rm tot} = 9.9\times 10^{12}M_\sun\left(10^{-3([{\rm O/H}]+1)}\right)
         \left(\frac{n_e}{10^{-4} {\rm\,cm}^{-3}}\right)^{-2} f 
\end{equation}
This is several times larger than estimates of the total Local Group binding
mass, e.g.\ $M_{\rm tot}= (2.3\pm 0.6)\times 10^{12} M_\sun$ as calculated by 
\citet{courteau99}.  Baryonic matter should only contribute $\sim 15\%$
of this mass (if the baryon--to--dark matter ratio is equal to the
cosmological value), so our estimate appears high.  This
discrepancy can be easily resolved with different values of [O/H] and $f$.
For instance, if we assume an oxygen abundance of $0.3$ times solar rather 
than 0.1, then the range of possible masses becomes 
$2.0\times 10^{11}f M_\sun \le M_{\rm tot} \le 7.9\times 10^{12}f M_\sun$.
Thus, unless [O/H] is very high or $f\ll 1$, the \ovii\ absorber almost 
certainly accounts for a major fraction of the baryonic matter in the 
Local Group.  This sightline may also be probing gas that is not
gravitationally bound to the Local Group (i.e., the Local Filament), which
may explain why our range of $M_{\rm tot}$ extends to such high values.
Although \citet{collins05} argue that a Local Group origin
requires that the \ovii\ absorbers contain too much mass, we see here that
the total mass is in fact consistent with expectations.

If this absorption does only trace the Local Group medium, then constraints 
on the extent of the absorber can be derived by assuming
$M_{\rm tot}=0.15 M_{\rm LG}\approx 3.5\times 10^{11}M_\sun$.  In this
case, $r^3 f=3M_{\rm tot}/(4\pi \times 1.4\mu n_e m_{\rm H})$.  Taking the
$2\sigma$ upper limit of $\log n_{\rm H}=-3.9$, the $2\sigma$ lower limit on 
$M_{\rm LG}=1.1\times 10^{12}M_\sun$, and assuming
$f=1$, we obtain a lower limit of $r>0.2$\,Mpc for the \ovii\ absorption.
Similarly, the upper limit on the radius is $r<0.6 f^{-1/3}$\,Mpc.  This
seems somewhat small compared to the actual size of the Local Group, but
once again is dependent on the geometry of the absorber.  A value of
$f\sim 0.1$ brings this upper limit more in line with the Mpc scales expected
in the Local Group; this may indicate that the Mkn~421 line of sight probes
an extended, filamentary WHIM distribution.  This calculation also assumes 
that the density of the Local Group medium is constant with radius, when
the actual density profile is more likely centrally concentrated.  These
measurements are also affected by the \ovi\ discrepancy: in case (b),
only the lower density limit of $\log(n_e)>-4.7$
(from the $2\sigma$ \ovi\ K$\alpha$ upper limit)
applies, so the upper radius and mass limits are still valid.
Nevertheless, the consistency with the expected Local Group parameters is
intriguing.
 

\subsection{Comparisons to Other Studies}

\citet{kravtsov02} used constrained simulations to study the properties of
the Local Supercluster region; sky
maps produced from this simulation (their Figure~5) show filamentary
structures near the Mkn~421 direction, possibly corresponding to the
observed absorption.  Additionally, they note that a Local Group medium
would exhibit a low Doppler parameter ($b\la 60$\,km\,s$^{-1}$) out
to distances of $\sim 7$\,Mpc, consistent with our \ovii\ measurement.

The inferred properties of the X-ray absorption along this line of sight
also appear similar to other observations:  toward
3C~273, for example, 
\citet{fang03} find $5.36<\log T<6.08$ and comparable \ovii\ 
column density; however, their inferred \ovii\ Doppler parameter is 
significantly higher than that toward Mkn~421: $b>100\kms$.  On the
other hand, the $z=0$ absorber toward PKS~$2155-304$ \citep{nicastro02}
exhibits a temperature consistent with the Mkn~421 absorber, yet
inferred density about an order of magnitude lower.  
Compared to the two intervening
filaments seen toward Mkn~421 \citep{nicastro05a}, the density of the
$z=0$ absorption agrees with the derived lower limits ($\log n_e\ga -5$
for both filaments), but the filaments appear to exhibit higher temperatures
than that derived for the local absorption.  These variations
along different lines of sight simply demonstrate the complex nature
of the absorption, and the diversity of temperature and density environments
produced in the structure formation process \citep[e.g.,][]{kravtsov02}.

The detection of X-ray absorption lines toward the Large Magellanic Cloud
binary LMC X-3 by \citet{wang05} presents an important consideration for
these results as well.  The measured \ovii\ and \neix\ equivalent widths
and upper limits on \oviii\ and \ovii$_{K\beta}$ are all consistent
with the same lines measured toward Mkn~421 (albeit with much larger 
statistical errors).  Although this detection provides evidence of 
a hot intervening absorber between the Galaxy and LMC, it certainly
does not rule out a primarily extragalactic origin for the Mkn~421
absorber.  Any absorption, either Galactic or extragalactic, is likely
to be inhomogeneous; thus, it's entirely plausible that the LMC X-3
sightline probes hot Galactic gas (perhaps enhanced by winds or outflows
from both the Galaxy and LMC), while the absorption toward Mkn~421 is
primarily due to low--density nearby WHIM gas.

\section{Summary and Future Work}

Through long--duration \chandra\ and \fuse\ observations of Mkn~421 in
outburst, we have obtained unprecedented measurements of a variety
of $z\sim 0$ absorption lines, many of which likely arise in extragalactic,
partially photoionized gas.  A brief summary of our results follows.
\begin{enumerate}
\item{The relative strengths of the three \ovii\ K-series lines imply
$2\sigma$ Doppler parameter constraints of $24<b<55$\,km\,s$^{-1}$.  This
is inconsistent with the value of $b=80.6\pm 4.2$\,km\,s$^{-1}$ 
derived for the Galactic low--velocity \ovi, indicating that the \ovilv\ 
and \ovii\ likely arise in different phases.  The \ovii\ $b$ value
is, however, consistent with the local IGM simulations of \citet{kravtsov02}
out to distances of several Mpc.}
\item{A weak high--velocity \ovi\ 1032\,\AA\ component also appears in the
\fuse\ spectrum.  Although its width is consistent with the \ovii\ $b$
measurement, the \ovi$_{\rm HVC}$ column density is too low to be associated
with the \ovii\ absorption unless $T\ga 10^7$\,K (which itself is ruled
out by the upper limit on $b$).  The \ovi$_{\rm HVC}$ may thus represent
a distinct third component along this line of sight.}
\item{The column density inferred from the \ovi\,K$_\alpha$ line is
a factor of $\sim 4$ higher than that measured from the \ovi\ 1032\,\AA\
transition.  This may be due to unaccounted--for atomic physics effects,
in which case the K$\alpha$ line may provide a more accurate measurement
of $\novi$ than the 1032\,\AA\ line.  We consider both cases:

  \begin{enumerate}
  \item{If the \ovi\ K$\alpha$ line measures $\novi$, then strong
  constraints on the
  temperature, density, and relative abundances of the X-ray absorber can 
  be derived:
  $T=10^{5.5-5.7}$\,K and $n_e=10^{-4.7}-10^{-3.9}$\,cm$^{-3}$, 
  with allowed abundances of $0.6\le$[Ne/O]$\le 2.2$, $-0.8\le$[C/O]$\le 0.3$,
  and [N/O]$\le 0.9$ (all $2\sigma$ ranges).  This range of densities, 
  combined with $\novii$, 
  implies a total mass and extent consistent with those expected in
  the Local Group and/or Local Filament if the gas metallicity is low.  
  However, in this case the
  \ovi\ UV--X-ray discrepancy becomes worse since (due to the Doppler
  parameter constraints) only a small portion of the \ovilv\ line can
  be associated with the extragalactic X-ray lines.}
  \item{If, instead, the \ovi\ K$\alpha$ line does not correctly measure
  $\novi$, then the \ovi\ associated with the \ovii\ absorption must
  be fully blended with the Galactic 1032\,\AA\ \ovi, and thus not measurable.
  In this case a lower density limit of $\log n_e>-4.7$ is found, which
  is consistent with either a Galactic or extragalactic medium.}
  \end{enumerate}
  }
\end{enumerate}

Much work remains to be done---both in order to better understand the
data presented here, and to determine the true nature of 
the $z\sim 0$ X-ray absorption.  Higher signal to noise data along the
Mkn~421 sightline would be useful to obtain better column density 
constraints, particularly on \oviii, \ovi$_{K\alpha}$, and \neix, and thus
better constrain the effects of photoionization on the absorber.  Data of
comparable quality along other sightlines would be invaluable as well,
both to probe other regions surrounding the Galaxy and to reconfirm
the tantalizing \ovi\ results presented herein.  Higher--resolution
simulations of the Local Group may allow us to determine the ionic
column densities expected in the WHIM, and thus whether or not the
observed absorption can possibly arise in the WHIM.  Finally, more 
detailed modeling of \ovi\ inner--shell transitions would shed
a great deal of light on whether or not the \ovi\ X-ray/UV discrepancy
is real, and thus provide an invaluable framework for studying new
(and existing) X-ray data.

\acknowledgements
We thank the \chandra\ and \fuse\ teams for superb missions, and the anonymous
referee for helpful suggestions during the review process.  R.~J.~Williams 
is supported by NASA through \chandra\ award AR5-6017X, 
F.~Nicastro through \chandra\ grant GO3-4152X (fund 16617317) 
and 5-year LTSA grant NNG04GD49G, and T.~Fang through {\sl
Chandra} Postdoctoral Fellowship Award Number PF3-40030.  \chandra\ awards
are issued by the
{\sl Chandra} X-ray Observatory Center, which is operated by the
Smithsonian Astrophysical Observatory for and on behalf of the NASA
under contract NAS8-39073.
This work is based on observations made with the NASA-CNES-CSA Far
Ultraviolet Spectroscopic Explorer.  FUSE is operated for NASA by the
Johns Hopkins University under NASA contract NAS5-32985.

\clearpage

\begin{deluxetable}{lcccrccc}
\tabletypesize{\footnotesize}
\tablecolumns{8}
\tablewidth{500pt}
\tablecaption{Observed $z\sim 0$ lines. \label{tab_lines}}
\tablehead{
\colhead{Line ID} &
\colhead{$\lambda_{\rm rest}$\tablenotemark{a}} &
\colhead{$\lambda_{\rm obs}$\tablenotemark{b}} &
\colhead{$\Delta v_{\rm FWHM}$} &
\colhead{$v_{\rm obs}$} &
\colhead{$W_\lambda$\tablenotemark{c}} &
\colhead{$\log N_i$\tablenotemark{c,d}} &
\colhead{Note} 
\\
\colhead{} &
\colhead{(\AA)} &
\colhead{(\AA)} &
\colhead{($\kms$)} &
\colhead{($\kms$)} &
\colhead{(m\AA)} &
\colhead{} &
\colhead{}
}

\startdata
X-ray: \\
\hline\hline
\ion{Ar}{15}$_{K\alpha}$ &24.737 & 24.737                  &\nodata & \nodata              & $<3.09$              & $<15.12$                & \\
\ion{C}{5}$_{K\alpha}$   &40.268 & $40.26\pm 0.02$         &\nodata & $-60\pm 150$         & $11.3^{+3.3}_{-2.6}$ & $15.19\pm 0.15$         & \\
\ion{C}{6}$_{K\alpha}$   &33.736 & $33.736\pm 0.02$        &\nodata & $0\pm 180$           & $7.2\pm 1.4$         & $15.31\pm 0.11$         & \\
\ion{Ne}{9}$_{K\alpha}$  &13.447 & $13.431\pm 0.02$        &\nodata & $-360\pm 450$        & $2.4^{+0.9}_{-0.8}$  & $15.48\pm 0.24$         & \\
\ion{Ne}{10}$_{K\alpha}$ &12.134 & $12.11^{+0.03}_{-0.02}$ &\nodata & $-590^{+740}_{-490}$ & $<5.04$              & $<16.21$                & 1\\
\ion{N}{6}$_{K\alpha}$   &28.787 & $28.755\pm 0.02$        &\nodata & $-330\pm 210$        & $4.1\pm 1.5$         & $15.02^{+0.19}_{-0.24}$ & \\
\ion{N}{7}$_{K\alpha}$   &24.781 & 24.781                  &\nodata & \nodata              & $<2.86$              & $<15.16$                & \\
\ion{O}{5}$_{K\alpha}$   &22.374 & 22.374                  &\nodata & \nodata             & $<2.20$         & $<14.97$         & 2,3\\
\ion{O}{6}$_{K\alpha}$   &22.019 & $22.023\pm 0.02$        &\nodata & $50\pm 270$          & $2.4\pm 0.9$         & $15.07^{+0.17}_{-0.22}$ & 2\\
\ion{O}{7}$_{K\alpha}$   &21.602 & $21.603\pm 0.02$        &\nodata & $10\pm 280$          & $9.4\pm 1.1$         & $16.22\pm 0.23$         & \\
\ion{O}{7}$_{K\beta}$    &18.629 & $18.612\pm 0.02$        &\nodata & $-273\pm 320$        & $4.6\pm 0.7$         & $16.28\pm 0.13$         & \\
\ion{O}{7}$_{K\gamma}$   &17.768 & $17.762\pm 0.02$        &\nodata & $-100\pm 340$        & $2.0\pm 0.7$         & $16.19^{+0.16}_{-0.21}$ & \\
\ion{O}{8}$_{K\alpha}$   &18.969 & $18.974\pm 0.02$        &\nodata & $80\pm 320$          & $1.8\pm 0.7$         & $15.17^{+0.16}_{-0.24}$ & \\
\hline
UV: \\
\hline\hline
\ion{O}{6}$_{1032}$             &1031.926 &$1031.88\pm 0.01$ &$152.3\pm 7.0$   &$-13.7\pm 2.6$ &$270.9\pm 7.9$ & $14.43\pm 0.02$         & \\
\ion{O}{6}$_{1032,\ {\rm HVC}}$ &1031.926 &$1032.30\pm 0.04$ &$59^{+31}_{-17}$ &$108\pm 12$    &$18.5\pm 5.6$  & $13.18^{+0.12}_{-0.16}$ & \\
\ion{O}{6}$_{1037}$             &1037.617 &$1037.59\pm 0.01$ &$145.4\pm 11.6$  &$-9.3\pm 4.0$  &$164.9\pm 8.4$ & $14.47\pm 0.03$         & \\
\enddata

\tablenotetext{a}{From \citet{verner96}, except \ovi\ and \ov\ which are from 
\citet{schmidt04}}
\tablenotetext{b}{Wavelength uncertainty is measured from fit or intrinsic
LETG 0.02\,\AA\ error, whichever is greater.  For upper limits, the line
position was frozen to the rest wavelength.}
\tablenotetext{c}{Error bars are $1\sigma$; upper limits are $2\sigma$.}
\tablenotetext{d}{Column densities are calculated using curve--of--growth
analysis with $b=80.6\pm 4.2\kms$ for \cv, \nvi, and \ov, and the \ovii\ 
Doppler parameter of $b=31-46\kms$ ($1\sigma$ range) for all other lines.}
\tablecomments{
(1) The \ion{Ne}{10} line lies within a detector feature, so only an upper
limit on its equivalent width is given. 
(2) The \ovi\ and \ov\ $\lambda_{\rm rest}$ values are from laboratory
measurements by \citet{schmidt04}; theoretical oscillator strengths
are taken from \citet{pradhan03}.
(3) There is an absorption line at $-760\kms$ from the \ov\ rest wavelength,
but this is more likely \ovii\ associated with Mkn~421.
}

\end{deluxetable}

\end{document}